\documentclass[a4paper,11pt]{article}
\usepackage[utf8]{inputenc}
\usepackage[dvips]{graphicx}

\usepackage{epsfig}
\usepackage{graphicx}
\usepackage{indentfirst}
\usepackage{amsmath}
\usepackage{amsfonts}
\usepackage{amssymb}

\begin{document}
\begin{center}
{\Large\bf From the early to the late time universe within $f(T,\mathcal{T})$ gravity}\\

\medskip

S. B. Nassur$^{(a)}$\footnote{e-mail:nassurmaeva@gmail.com}, M. J. S. Houndjo$^{(a,b)}$\footnote{e-mail:
sthoundjo@yahoo.fr}, A. V. Kpadonou$^{(a,c)}$\footnote{e-mail: vkpadonou@gmail.com}, M. E. Rodrigues$^{(d)}$\footnote{e-mail: esialg@gmail.com},  
 \\and J. Tossa$^{(a)}$\footnote{e-mail: joel.tossa@imsp-uac.org}

$^a$ \,{\it Institut de Math\'{e}matiques et de Sciences Physiques (IMSP)}\\
 {\it 01 BP 613,  Porto-Novo, B\'{e}nin}\\

$^{b}$\,{\it Facult\'e des Sciences et Techniques de Natitingou - Universit\'e de Parakou - B\'enin} \\

$^{c}$\, {\it Ecole Normale Sup\'erieure de Natitingou - Universit\'e de Parakou - B\'enin}\\ 

$^{d}$\,{Faculdade de Ci\^encias Exatas e Tecnologia, Universidade Federal do Par\'a - Campus Universit\'ario de
Abaetetuba, CEP 68440-000, Abaetetuba, Par\'a, Brazil}\\

%\date{}

\end{center}
\begin{abstract}
In this paper we perform the reconstruction scheme of the gravitational action within $f(T,\mathcal{T})$ gravity, where $T$ and $\mathcal{T}$ denote the torsion scalar and the trace of the energy momentum tensor, respectively. We particularly focus our attention on the case where the algebraic function $f(T,\mathcal{T})$ is decomposed as a sum of two functions $f_1(T)$ and $f_2(\mathcal{T})$, i.e, \newline $f(T,\mathcal{T})=f_{1}(T)+f_{2}(\mathcal{T})$. The description is essentially based on the scale factor and then, we consider two interesting and realistic expressions of this parameter and reconstruct the action corresponding to each phase of the universe. Our results show that some $f(T,\mathcal{T})$ models are able  to describe the evolution of the universe from the inflation phase to the late time dark energy dominated phase.
\end{abstract}

\section{Introduction}
It is well known that our universe is now experiencing an accelerated expansion. Several cosmological observational data defend this phenomenon as supernovae type Ia, cosmic microwave background radiation, large scale structure, baryon acoustic oscillations and weak lensing \cite{1001}\cite{1de1207.1646}\cite{2de1207.1646}. There are two representative approaches to explain this acceleration. The first is that the universe is filled by an exotic fluid with negative pressure called dark energy generally materialised by the cosmological constant within the General Relativity (GR). The second way is modifying the gravitational action and so, explanations can be done about the acceleration of the expansion of the universe. Several authors carried out some important and interesting results on this way within theories essential based on the curvature scalar \cite{1311},\cite{1202}\cite{1005}\cite{1011'}\cite{1205}.\par 
However, there is another modified theory of gravity, which, instead of the curvature, is essential based on the torsion of the spacetime through the Weitzenbock connection, called $f(T)$ theory of gravity. Various interesting results have been found within this theory \cite{1202}\cite{1011} \cite{bamba}\cite{20a57deines1}, but do not take into account the aspect where the cosmological constant may be variable, and more precisely dependent on the trace $\mathcal{T}$ of the energy-momentum tensor 
\cite{99}\cite{1004}\cite{1308}. Therefore, introducing $\mathcal{T}$ in the action gives rise to the $f(T, \mathcal{T})$ theory of gravity. This theory has been performed first by \cite{saridakis}\cite{1405} with interesting results.  Note that this kind of theory has been developed  through the consideration of curvature, called $f(R,\mathcal{T})$ theory of gravity where $R$ denotes the curvature of the spacetime, and potential results have been obtained \cite{prof}\cite{1308}.\par
In this paper, we focus our attention on the $f(T,\mathcal{T})$ theory of gravity, which is a generalisation of $f(T)$ theory.
As it is well known through the cosmological observational data, our universe had been dominated by the matter where the expansion was decelerated and later, entered in the actual phase, dominated by the so-called dark energy, where the expansion is accelerated. Our goal in this paper is to describe this cosmological evolution of the universe, within some suitable expressions of the scale factor, by characterising the algebraic functions of the gravitational action associated to each phase, using the so-called cosmological reconstruction scheme. More precisely, two realistic expressions have been considered for the scale factor and our  results show that there exists $f(T,\mathcal{T})$ models able to describe the two important phases of the evolution of the universe and the transition from the matter dominated phase to the dark energy dominated one. \par
The paper is organised as follows. The Section $2$ points out the generality on the $f(T,\mathcal{T})$, where the field equations have been carried out in the framework of Friedmann-Robertson-Walker metric. The Section $3$ addresses the unification of the matter dominated and the dark energy dominated phases. The transition between the matter dominated and dark energy dominated phases is performed in the Section $4$. We undertake a second approach, view as a more general case of the previous one, where the results obtained in the previous Sections have been reobtained for very special limit of some input parameters. Finally the conclusion is presented in the Section $6$.

\section{Generality}
Let us consider the action $S$ of the $f(T,\mathcal{T})$ gravity, given by 
\begin{eqnarray}
\label{1}
 S=\int d^{4}xe\left[f(T,\mathcal{T})+\mathcal{L}_{m}\right],
\end{eqnarray}

where $\mathcal{T}=\delta_{\mu}^{\nu}\mathcal{T}_{\nu}\,^{\mu}$ is the trace of the energy-momentum tensor and $\mathcal{L}_{m}$ the matter Lagrangian density assumed to depend only on the tetrad, and not on its covariant derivatives. $e$ is the determinant of the tetrad and $f$ an arbitrary function of the scalar torsion  $T$ and the trace of the energy-momentum tensor  $\mathcal{T}$. Here  we set $16\pi G=1$.\par 
The variation of the action (\ref{1}) with respect to the tetrad leads to  \cite{saridakis}
\begin{eqnarray}
\label{2}
 \left[e^{-1}\partial_{\mu}\left(ee_{a}\,^{\alpha}S_{\alpha}\,^{\rho\mu}\right)-e_{a}\,^{\alpha}T^{\mu}\,_{\nu\alpha}S_{\mu}\,^{\nu\rho}\right]f_{T}
 +e_{a}\,^{\alpha}S_{\alpha}\,^{\rho\mu}\left(f_{TT}\partial_{\mu}T+f_{T\mathcal{T}}\partial_{\mu}\mathcal{T}\right)+\frac{e_{a}\,^{\rho}f}{4}\nonumber\\
 -\left(\frac{e_{a}\,^{\alpha}\mathcal{T}_{\alpha}\,^{\rho}+pe_{a}\,^{\rho}}{2}\right)f_{\mathcal{T}}=\frac{1}{4}e_{a}\,^{\alpha}\mathcal{T}_{\alpha}\,^{\rho}.
\end{eqnarray}
with $f_{TT}=\partial^{2}f/(\partial_{T})^{2}$, $f_{T\mathcal{T}}=\partial^{2}f/\partial_{T}\partial_{\mathcal{T}}$, $f_{T}=\partial f/\partial_{T}$ et $f_{\mathcal{T}}=\partial f/\partial_{\mathcal{T}}$.
 \\
One can equivalently write the action (\ref{1}) on the following form \cite{mannuel}\cite{0904}\cite{prof}
\begin{eqnarray}
\label{a}
S=\int d^{4}xe\left[P_{1}(\Phi)T+P_{2}(\Phi)\mathcal{T}+Q(\Phi)+\mathcal{L}_{m}\right],
\end{eqnarray}
where $P_{1}, P_{2}$ and  $Q$ are proper functions of the scalar field $\Phi$. One supposes that the field  $\Phi$ does not have kinetic term and may be interpreted as an auxiliary field. Therefore, we can variate the action (\ref{a}) with respect to this auxiliary field $\Phi$, getting 
\begin{eqnarray}
\label{b}
 P'_{1}(\Phi)T+P'_{2}(\Phi)\mathcal{T}+Q'(\Phi)=0.
\end{eqnarray}
In the view that this equation is solvable, its solution can be written as function of both the torsion scalar and the trace of the energy-momentum tensor as 
\begin{eqnarray}
 \label{c}
 \Phi=\Phi(T,\mathcal{T}).
\end{eqnarray}
Here we use the signature  $(+,-,-,-)$, and assume that the universe is flat and homogeneous for large scales. Then, we use the flat FRW metric and get \cite{1405}
\begin{eqnarray}
\label{3}
 e^{a}\,_{\mu}=diag[1,a,a,a],\quad T^{j}\,_{0i}=H\delta_{i}^{j},\quad S_{j}\,^{0i}=-H\delta_{j}^{i},
\end{eqnarray}
and the torsion scalar reads 
\begin{eqnarray}
\label{4}
 T=-6H^{2}.
\end{eqnarray}
By using the relations  (\ref{3}) and  (\ref{4}) in the equation of motion  (\ref{2}), one gets \cite{saridakis}
\begin{eqnarray}
\label{5}
 3H^{2}f_{T}+\frac{f}{4}=\frac{\rho}{4}+\frac{\rho+p}{2}f_{\mathcal{T}},\\
 \label{6}
 -12H^{2}\dot{H}f_{TT}+(3H^{2}+\dot{H})f_{T}+\frac{f}{4}=-\frac{p}{4}-H(\dot{\rho}-3\dot{p})f_{T\mathcal{T}}.
\end{eqnarray}
 \subsection*{  $\bullet$  Decoupling the gravitational Lagrangian density}
Let us consider  that the function $f$ is the sum of two independent functions $f_{1}$ and $f_{2}$, respectively depending on the torsion scalar $T$ and the trace $\mathcal{T}$, i.e, 
\begin{eqnarray}
\label{7}
 f(T,\mathcal{T})=f_{1}(T)+f_{2}(\mathcal{T}),
\end{eqnarray}
with 
\begin{eqnarray}
\label{8}
 f_{1}(T)=P_{1}(\Phi)T+Q_{1}(\Phi),\quad f_{2}(\mathcal{T})=P_{2}(\Phi)\mathcal{T}+Q_{2}(\Phi),\quad Q(\Phi)=Q_{1}(\Phi)+Q_{2}(\Phi).
\end{eqnarray}
By using the relations (\ref{7}) and  (\ref{8}), one gets 
\begin{eqnarray}
\label{9}
 f_{T}(T,\mathcal{T})&=&P_{1}(\Phi)+\Phi_{T}\left[P'_{1}(\Phi)T+Q'_{1}(\Phi)+Q'_{2}(\Phi)+P'_{2}(\phi)\mathcal{T}\right],\nonumber\\
 &=&P_{1}(\Phi),\nonumber\\
 f_{TT}(T,\mathcal{T})&=&P_{1T}(\Phi)=\Phi_{T}P'_{1}(\Phi),\nonumber\\
 f_{\mathcal{T}}(T,\mathcal{T})&=&P_{2}(\Phi)+\Phi_{\mathcal{T}}\left[P'_{1}(\Phi)T+Q'_{1}(\Phi)+Q'_{2}(\Phi)+P'_{2}(\phi)\mathcal{T}\right],\nonumber\\
 &=&P_{2}(\Phi),\\
 %f_{T\mathcal{T}}(T,\mathcal{T})&=&P_{1\mathcal{T}}(\Phi)=P_{2T}(\Phi)=\Phi_{T}P'_{2}(\Phi).\nonumber\\
 f_{T\mathcal{T}}(T,\mathcal{T})&=&\Phi_{\mathcal{T}}P'_{1}(\Phi)=\Phi_{T}P'_{2}(\Phi).\nonumber
\end{eqnarray}
where $\Phi_T$ and $\Phi_{\mathcal{T}}$ are the derivatives with respect to $T$ and $\mathcal{T}$ respectively. By suting the relations  (\ref{7}), (\ref{8}) and  (\ref{9}) into (\ref{5}) and  (\ref{6}), one gets 
\begin{eqnarray}
 \label{10}
 6H^{2}P_{1}(\Phi)+Q(\Phi)&=&\rho+(\rho+5p)P_{2}(\Phi),\\
 \label{11}
8H\dot{P}_{1}(\Phi)+(6H^{2}+4\dot{H})P_{1}(\Phi)+Q(\Phi)&=&-p-(\rho-3p)P_{2}(\Phi).
\end{eqnarray}
Here the ``dot" denotes the derivative with respect to the time $t$. We have two equations with three unknown variables  $P_{1}(\Phi),P_{2}(\Phi)$ and $Q(\Phi)$. 
\par 
From the relations (\ref{10}) and (\ref{11}), we define the effective energy density  and pressure  $\rho_{eff}$ and $p_{eff}$ as follows
\begin{eqnarray}
\label{13}
 \rho_{eff}&=&\rho+(\rho+5p)P_{2}(\Phi)-6H^{2}\left(P_{1}(\Phi)-1\right)-Q(\Phi),\\
 \label{14}
 p_{eff}&=&p+(\rho-3p)P_{2}(\Phi)+(6H^{2}+4\dot{H})\left(P_{1}(\Phi)-1\right)+Q(\Phi)+8H\dot{P}_{1}(\Phi).
\end{eqnarray}
Note that by setting $f(T,\mathcal{T})=T, P_{1}(\Phi)=0, P_{2}(\Phi)=0$ and $Q(\Phi)=0$, the effective pressure $p_{eff}$ coincide with $p$,and $\rho_{eff}$ with $\rho$. Then, the relations  (\ref{13}) and (\ref{14}) are the modified equations of Friedmann. In this theory, the effective energy density  and the ordinary one are conserved separately, i.e,
\begin{eqnarray}
\label{c1}
 \dot{\rho}_{eff}+3H(\rho_{eff}+p_{eff})=0,\\
 \label{c2}
 \dot{\rho}+3H(\rho+p)=0.
\end{eqnarray}
Making use of the barotropic relation  $p=\omega\rho$, one gets 
\begin{eqnarray}
\label{15}
 18H^{2}\dot{P}_{1}(\Phi)-\dot{Q}(\Phi)+\rho(1+5\omega)\dot{P}_{2}(\Phi)-3H\rho(5\omega^{2}+4\omega-1)P_{2}(\Phi)=0.
\end{eqnarray}
The equation (\ref{15})  can be regarded as an equation describing the evolution of the Hubble parameter $H$ and the energy density $\rho$ in an homogeneous and isotropic universe, which in this case does not depend on spatial coordinates. Therefore, the auxiliary field may be considered as the cosmic time, i.e, 
\begin{eqnarray}
 \Phi=t.
\end{eqnarray}
On the other hand one can eliminate $Q(\Phi)$ by subtracting  (\ref{10}) from  (\ref{11}) in order to obtain a relation between  $P_{1}(\Phi)$ and $P_{2}(\Phi)$, 
\begin{eqnarray}
\label{12}
8H\dot{P}_{1}(\Phi)+4\dot{H}P_{1}(\Phi)&=&-(\rho+p)(1+2P_{2}(\Phi)).
\end{eqnarray}
In order to obtain the algebraic function  $f$, one chooses a cosmological scale factor $a(t)$. To this end, we use the exponential evolution of the scale factor as 
\begin{eqnarray}
\label{19}
 a(t)=a_{0}e^{g(t)},
\end{eqnarray}
where $g(t)$ is a non-linear function depending on the time, and $a_{0}$ a positive constant. The equation (\ref{12}) becomes
\begin{eqnarray}
\label{16}
8\dot{g}\frac{dP_{1}}{d\Phi}(\Phi)+4\ddot{g}P_{1}(\Phi)=-\rho(1+\omega)(1+2P_{2}(\Phi)).
\end{eqnarray}
By fixing $P_{2}(\phi)$, one determines  $P_{1}(\Phi)$ through the resolution of the equation (\ref{16}) and using the relation  (\ref{10}), one finds  $Q(\phi)$.
\section{ Unification of the matter dominated and accelerated dark energy dominated phases}
The simple resolution of the equation (\ref{16}) can be done by vanishing the right side, then, setting  $P_{2}(\Phi)=-\frac{1}{2}$ and considering the following example  \cite{prof}
%%%%%%%%%%%%%%%%%%%%%%%%%%%%%%%%%%%%%%%%%%%%%%%%%%%%%%%%%%%%%%%%%%%%%%%%%%%%%%%%%%%%%%%%%%
%%%%%%%%%%%%%%%%%%%%%%%%%%%%%%%%%%%%%%%%%%%%%%%%%%%%%%%%%%%%%%%%%%%%%%%%%%%%%%%%%%%%%%%%%
%%%%%%%%%%%%%%%%%%%%%%%%%%%%%%%%%%%%%%%%%%%%%%%%%%%%%%%%%%%%%%%%%%%%%%%%%%%%%%%%%%%%%%%%%
%%%%%%%%%%%%%%%%%%%%%%%%%%%%%%%%%%%%%%%%%%%%%%%%%%%%%%%%%%%%%%%%%%%%%%%%%%%%%%%%%%%%%%%%%
%%%%%%%%%%%%%%%%%%%%%%%%%%%%%%%%%%%%%%%%%%%%%%%%%%%%%%%%%%%%%%%%%%%%%%%%%%%%%%%%%%%%%%%%%
\begin{eqnarray}
\label{18}
 g(\Phi)=g_{0}\Phi+g_{1}\ln(\Phi),
\end{eqnarray}
with $g_{0}$ and $g_{1}$ two positive constants. Thereby one gets
\begin{eqnarray}
\label{20}
\dot{g}\equiv H=g_{0}+\frac{g_{1}}{\Phi},\ddot{g}\equiv \dot{H}=-\frac{g_{1}}{\Phi^{2}}, a=a_{0}e^{g_{0}\Phi}\Phi^{g_{1}}. % H(t)=g_{0}+\frac{g_{1}}{t}.
\end{eqnarray}
From the relations (\ref{16}) and (\ref{10}), one obtains 
\begin{eqnarray}
\label{23}
 P_{1}(\Phi)=C\sqrt{\frac{\Phi}{g_{0}\Phi+g_{1}}}=\frac{C}{g_{0}^{1/2}}\left(\frac{1}{1+\frac{g_{1}}{g_{0}\Phi}}\right)^{1/2},\\
 Q(\Phi)=\frac{\rho_{0}(1-5\omega)}{2a_{0}^{3(1+\omega)}e^{3g_{0}(1+\omega)\Phi}\Phi^{3g_{1}(1+\omega)}}-6Cg_{0}^{3/2}\left(1+\frac{g_{1}}{g_{0}\Phi}\right)^{3/2},
 \end{eqnarray}
with $C$  an integration constant.  Now, we can determine the expression of $\Phi$ from the equation (\ref{b}). However, we can take an asymptotic approach through a simplification. 
\begin{enumerate}
 \item $0<\Phi\ll1$ 
The equation (\ref{b}) can be written as
\begin{eqnarray}
  9Cg_{1}^{3/2}\Phi^{-2}+\frac{9g_{0}g_{1}\rho_{0}(1-5\omega)(1+\omega)^{2}\Phi^{3g_{1}(1+\omega)+1/2}}{2a_{0}^{-3g_{1}(1+\omega)}}\nonumber\\
  -\frac{3\rho_{0}(1-5\omega)(1+\omega)g_{1}\Phi^{-(3g_{1}(1+\omega)+1/2)}}{2a_{0}^{3(1+\omega)}}+\frac{CT}{g_{1}^{1/2}}\simeq0.
 \end{eqnarray}
We can distinguish two cases according to the exponent of $\Phi$.
\begin{itemize}
 \item If  $g_{1}(1+\omega)>\frac{1}{2}$ then
 \begin{eqnarray}
  \Phi&\sim& \alpha_{1}T^{-\alpha_{0}},\\
  \alpha_{0}&\equiv&\frac{2}{6g_{1}(1+\omega)+1},\nonumber\\
  \alpha_{1}&\equiv&\left(\frac{3g_{0}^{3/2}\rho_{0}(1-5\omega)(1+\omega)}{2Ca_{0}^{3(1+\omega)}}\right)^{\alpha_{0}}.\nonumber
 \end{eqnarray}
Then, one obtains
\begin{eqnarray}
 f(T,\mathcal{T})&\simeq&\frac{CT}{g_{0}^{1/2}}\left(1+\frac{g_{1}T^{\alpha_{0}}}{g_{0}\alpha_{1}}\right)^{-1/2}-6Cg_{0}^{3/2}\left(1+\frac{g_{1}T^{\alpha_{0}}}{g_{0}\alpha_{1}}\right)^{3/2}\nonumber\\
 &&+\frac{\rho_{0}(1-5\omega)T^{3\alpha_{0}g_{1}(1+\omega)}}{2a_{0}^{3(1+\omega)}e^{3g_{0}\alpha_{1}(1+\omega)T^{-\alpha_{0}}}\alpha_{1}^{3g_{1}(1+\omega)}}-\frac{1}{2}\mathcal{T}. 
\end{eqnarray}
\item If $g_{1}(1+\omega)<\frac{1}{2}$, then 
\begin{eqnarray}
 \Phi\sim 3g_{1}\left(-\frac{1}{T}\right)^{1/2}.
\end{eqnarray}
Therefore one has 
\begin{eqnarray}
 f(T,\mathcal{T})&\simeq&\frac{CT}{g_{0}^{1/2}}\left(1+\frac{\sqrt{-T}}{3g_{0}}\right)^{-1/2}-6Cg_{0}^{3/2}\left(1+\frac{\sqrt{-T}}{3g_{0}}\right)^{3/2}\nonumber\\
 &&+\frac{\rho_{0}(1-5\omega)(-T)^{3g_{1}(1+\omega)/2}}{2(3g_{1})^{3g_{1}(1+\omega)}a_{0}^{3(1+\omega)}e^{9g_{0}(1+\omega)g_{1}\sqrt{-1/T}}}-\frac{1}{2}\mathcal{T}.
\end{eqnarray}
Note that we use the same initial values as in \cite{prof}. This leads to a Lagrangian density easy to be analysed because the exponent of  $T$ is fixed and this is proper to $f(T,\mathcal{T})$ gravity, in opposite to the $f(R,\mathcal{T})$ gravity, where the Lagrangian density obtained within these conditions are complex \cite{prof}.
\item $\Phi\longrightarrow+\infty.$\\
In this limit, the relation (\ref{b}) becomes
\begin{eqnarray}
 \left(\frac{9g_{1}^{2}}{2g_{0}^{1/2}}-\frac{3CTg_{1}^{2}}{4g_{0}^{2}}\right)\frac{1}{\Phi}+\frac{CTg_{1}}{2g_{0}^{3/2}}+9g_{1}g_{0}^{1/2}\sim0,
\end{eqnarray}
leading to 
\begin{eqnarray}
 \Phi\sim\frac{3g_{1}(6g_{0}^{1/2}-CT)}{2g_{0}^{1/2}(-CT-18g_{0}^{2})},
\end{eqnarray}
with $-CT>18g_{0}^{2}.$
Therefore  
%\begin{eqnarray}
% f(T,\mathcal{T})&\simeq&\frac{CT}{g_{0}^{1/2}}\left(1-\frac{2(CT+18g_{0}^{2})}{3g_{0}^{1/2}(6g_{0}^{1/2}-CT)}\right)^{-1/2}-\frac{1}{2}\mathcal{T}\nonumber\\
% &&+\left(\frac{2\sqrt{g_{0}}}{3g_{1}}\right)^{3g_{1}(1+\omega)}\frac{\rho_{0}(1-5\omega)}{2a_{0}^{3(1+\omega)}}\left(\frac{-CT-18g_{0}^{2}}{6g_{0}^{1/2}-CT}\right)^{3g_{1}(1+\omega)}\nonumber\\
% &&Exp\left(-\frac{9g_{1}g_{0}^{1/2}(1+\omega)(6g_{0}^{1/2}-CT)}{2(-CT-18g_{0}^{2})}\right)\nonumber\\
% &&-6Cg_{0}^{3/2}\left(1-\frac{2(CT+18g_{0}^{2})}{3g_{0}^{1/2}(6g_{0}^{1/2}-CT)}\right)^{3/2}.
%\end{eqnarray}
\begin{eqnarray}
 f(T,\mathcal{T})=\frac{CT}{\sqrt{g_{0}\Psi_{0}(T)}}-6C\sqrt{(g_{0}\Psi_{0}(T))^{3}}+\Psi_{1}(T)-\frac{1}{2}\mathcal{T},
\end{eqnarray}
with 
\begin{eqnarray}
 \Psi_{0}(T)&\equiv&\left(1-\frac{2(CT+18g_{0}^{2})}{3g_{0}^{1/2}(6g_{0}^{1/2}-CT)}\right),\\
 \Psi_{1}(T)&\equiv&\left(\frac{2\sqrt{g_{0}}}{3g_{1}}\right)^{3g_{1}(1+\omega)}\frac{\rho_{0}(1-5\omega)}{2a_{0}^{3(1+\omega)}}\left(\frac{-CT-18g_{0}^{2}}{6g_{0}^{1/2}-CT}\right)^{3g_{1}(1+\omega)}\\
 &&\times e^{-\frac{9g_{1}g_{0}^{1/2}(1+\omega)(6g_{0}^{1/2}-CT)}{2(-CT-18g_{0}^{2})}}.
\end{eqnarray}

\end{itemize}

\end{enumerate}

\section{Transition between the matter dominated and dark energy dominated phases}
In this section, we will consider $g(\Phi)$ on the form 
\begin{eqnarray}
\label{24}
 g(\Phi)=h(\Phi)\ln(\Phi),
\end{eqnarray}
with $h(\Phi)$ a function changing very slowly as the field $\Phi$ evolves. Making use of the definition  $(\ref{24})$ for an adiabatic approximation of  $h(\phi)$ i.e, $(h'\sim h''\sim0)$ in the equation (\ref{16}),
one gets
\begin{eqnarray}
\label{23'}
 P_{1}(\Phi)=P_{0}\sqrt{\Phi},
\end{eqnarray}
with $P_{0}$ an integration constant, with $P_{2}=-\frac{1}{2}.$ 
%one obtains a solution on the form  (\ref{23}) i.e $P_{1}(\Phi)=C\sqrt{\Phi}$ 
%if we take $P_{2}(\phi)=-\frac{1}{2}$.\\
On the other hand the definition (\ref{24}) yields 
\begin{eqnarray}
\label{26}
 H(t)\sim\frac{h(t)}{t},\quad T\sim-\frac{6h^{2}(t)}{t^{2}}.
\end{eqnarray}
Let us take $h(\Phi)$ on the form:
\begin{eqnarray}
\label{25}
 h(\Phi)=\frac{h_{i}+h_{f}q\Phi^{2}}{1+q\Phi^{2}},
\end{eqnarray}
with $q, h_{i}$ and $h_{f}$ positive constants. From the definition (\ref{25}), one gets 
\begin{eqnarray}
\label{27}
 \lim_{\Phi\longrightarrow0}h(\Phi)=h_{i},\quad\lim_{\Phi\longrightarrow\infty}h(\Phi)=h_{f},
\end{eqnarray}
and  
\begin{eqnarray}
\label{acc}
 \frac{\ddot{a}}{a}=\frac{h(\Phi)}{\Phi^{2}}\left[-1+h(\Phi)\right].
\end{eqnarray}
Taking  $0<h_{i}<1$, the pass is characterized by a decelerated expansion of the universe, while for  $h_{f}>1$, the future is characterized by an accelerated expansion of the universe. By considering the case where  $h_{i}>1, 0<h_{m}<1$ and  $h_{f}>1$, one gets an accelerated universe which should be interpreted as the inflationary phase, followed by the decelerated phase corresponding a matter dominated phase and finally an accelerated dark energy dominated phase. We note that  $h_{m}$ is the value of $h(t)$ when $0\ll t_{m}\ll+\infty$.\\
We note from (\ref{26}) and (\ref{27}) that when $\Phi$ is low, i.e  $T\sim-6\frac{h^{2}_{i}}{\Phi^{2}}$ is high (absolute value), and contrary, $T\sim-6\frac{h^{2}_{f}}{\Phi^{2}}$ is low  
(absolute value) when $\Phi$ is large. An asymptotic approach can be undertaken for determining $f(T,\mathcal{T})$ able to reproduce the transition and also the important cosmological epochs of the universe with the definition (\ref{24}).\\ 
By using the relation (\ref{10}), it appears that 
\begin{eqnarray}
 Q(\Phi)=\frac{1}{2}\rho_{0}a_{0}^{-3(1+\omega)}(1-5\omega)\Phi^{-3h(\Phi)(1+\omega)}-6Ch^{2}(\Phi)\Phi^{-\frac{3}{2}}.
\end{eqnarray}

When $\Phi$ is low, one gets 
\begin{eqnarray}
 Q(\Phi)_{i}\simeq\frac{1}{2}\rho_{0}a_{0}^{-3(1+\omega)}(1-5\omega)\Phi^{-3h_{i}(1+\omega)}-6Ch^{2}_{i}\Phi^{-\frac{3}{2}}.
\end{eqnarray}
Then the relation (\ref{b}) becomes
\begin{eqnarray}
 CT-3h_{i}\rho_{0}a_{0}^{-3(1+\omega)}(1+\omega)(1-5\omega)\Phi^{-3h_{i}(1+\omega)-1/2}+18Ch_{i}^{2}\Phi^{-2}\simeq0.
\end{eqnarray}
It comes that 
\begin{itemize}
 \item For $h_{i}(1+\omega)>1/2\Rightarrow h_{i}>1/2(1+\omega)$, from the relation (\ref{acc}), there is acceleration with  $(h_{i}>1)$ and a deceleration with $1>h_{i}>1/2(1+\omega)$. Therefore, the following action can describe the transition between the inflation phase and the radiation phase. Then 

 \begin{eqnarray}
  \Phi\sim\beta_{1}\left(-\frac{1}{T}\right)^{\beta_{0}},
 \end{eqnarray}
with 
\begin{eqnarray}
 \beta_{0}\equiv\frac{2}{3h_{i}(1+\omega)+1}, \beta_{1}\equiv\left(\frac{3h_{i}\rho_{0}(5\omega-1)(1+\omega)}{Ca_{0}^{3(1+\omega)}}\right)^{\beta_{0}}
\end{eqnarray}
which leads to 
\begin{eqnarray}
\label{inf-rad}
% f(T,\mathcal{T})\simeq C\sqrt{\beta_{1}}T(-T)^{-\beta_{0}/2}+\frac{\rho_{0}(1-5\omega)(-T)^{3\beta_{0}h_{i}(1+\omega)}}{2a_{0}^{3(1+\omega)}\beta_{1}^{3h_{i}(1+\omega)}}-\frac{6Ch^{2}_{i}(-T)^{3\beta_{0}/2}}{\beta_{1}^{3/2}}-\frac{\mathcal{T}}{2}.\nonumber\\
 f(T,\mathcal{T})\simeq C_{1}T(-T)^{-\beta_{0}/2}+C_{2}(-T)^{\beta_{2}}+C_{3}(-T)^{3\beta_{0}/2}-\frac{1}{2}\mathcal{T},
\end{eqnarray}
where 
\begin{eqnarray}
 C_{1}\equiv C\sqrt{\beta_{1}}, C_{2}\equiv\frac{\rho_{0}(1-5\omega)}{2a_{0}^{3(1+\omega)}\beta_{1}^{3h_{i}(1+\omega)}},C_{3}\equiv-\frac{6Ch^{2}_{i}}{\beta_{1}^{3/2}},\beta_{2}\equiv3\beta_{0}h_{i}(1+\omega).
\end{eqnarray}

\item For $h_{i}(1+\omega)<1/2\Rightarrow h_{i}<1/2(1+\omega)$, the action can describe a decelerated universe which can be interpreted as radiation phase. Then
\begin{eqnarray}
 \Phi\sim3h_{i}\sqrt{-\frac{2}{T}}.
\end{eqnarray}
Consequently one gets 
\begin{eqnarray}
\label{rad}
% f(T,\mathcal{T})\simeq-C(18h^{2}_{i})^{1/4}(-T)^{3/4}+\frac{\rho_{0}(1-5\omega)}{2a_{0}^{3(1+\omega)}(18h^{2}_{i})^{3h_{i}(1+\omega)/4}}(-T)^{3h_{i}(1+\omega)/2}-\frac{6Ch^{2}_{i}}{(18h^{2}_{i})^{3/8}}(-T)^{3/4}-\frac{\mathcal{T}}{2},\nonumber\\
%  f(T,\mathcal{T})\simeq-C\left((18h^{2}_{i})^{1/4}+\frac{6h^{2}_{i}}{(18h^{2}_{i})^{3/8}}\right)(-T)^{3/4}+\frac{\rho_{0}(1-5\omega)}{2a_{0}^{3(1+\omega)}(18h^{2}_{i})^{3h_{i}(1+\omega)/4}}(-T)^{3h_{i}(1+\omega)/2}-\frac{\mathcal{T}}{2}.\\
  f(T,\mathcal{T})\simeq C'_{1}(-T)^{3/4}+C'_{2}(-T)^{3h_{i}(1+\omega)/2}-\frac{\mathcal{T}}{2},
\end{eqnarray}
with
\begin{eqnarray}
 C'_{1}\equiv-C\left((18h^{2}_{i})^{1/4}+\frac{6h^{2}_{i}}{(18h^{2}_{i})^{3/8}}\right), C'_{2}\equiv\frac{\rho_{0}(1-5\omega)}{2a_{0}^{3(1+\omega)}(18h^{2}_{i})^{3h_{i}(1+\omega)/4}}.
\end{eqnarray}
\end{itemize}
%%%%%%%%%%%%%%%%%%%%%%%%%%%%%%%%%%%%%%%%%%%%%%%%%%%%%%%%%%%%%%%%%%%%%%%%%%%%%%%%%%%%%%%%%%%%%%%%%%%%%%%%%%%%%%%%%%%%%%%%%%%%%%%%%%%%%%%%%%%%%%%%%%%%%%%
%%%%%%%%%%%%%%%%%%%%%%%%%%%%%%%%%%%%%%%%%%%%%%%%%%%%%%%%%%%%%%%%%%%%%%%%%%%%%%%%%%%%%%%%%%%%%%%%%%%%%%%%%%%%%%%%%%%%%%%%%%%%%%%%%%%%%%%%%%%%%%%%%%%%%%%
When $\Phi$ is large, one gets  
\begin{eqnarray}
 Q(\Phi)_{f}\simeq\frac{1}{2}\rho_{0}a_{0}^{-3(1+\omega)}(1-5\omega)\Phi^{-3h_{f}(1+\omega)}-6Ch^{2}_{f}\Phi^{-\frac{3}{2}}.
\end{eqnarray}
From the relation (\ref{b}), one gets 
\begin{eqnarray}
 CT-3h_{f}\rho_{0}a_{0}^{-3(1+\omega)}(1+\omega)(1-5\omega)\Phi^{-3h_{f}(1+\omega)-1/2}+18Ch_{f}^{2}\Phi^{-2}\simeq0.
\end{eqnarray}
Thus one can distinguish the following points
\begin{itemize}
\item for $h_{f}<\frac{1}{2(1+\omega)}$, one has
 \begin{eqnarray}
  \Phi\sim\beta_{4}\left(-\frac{1}{T}\right)^{\beta_{3}},
 \end{eqnarray}
 with
\begin{eqnarray}
 \beta_{3}\equiv\frac{2}{3h_{f}(1+\omega)+1}, \beta_{4}\equiv\left(\frac{3h_{f}\rho_{0}(5\omega-1)(1+\omega)}{Ca_{0}^{3(1+\omega)}}\right)^{\beta_{3}},
\end{eqnarray}
leading to
\begin{eqnarray}
\label{ord}
%f(T,\mathcal{T})\simeq C\sqrt{\beta_{4}}T(-T)^{-\beta_{3}/2}+\frac{\rho_{0}(1-5\omega)(-T)^{3\beta_{3}h_{f}(1+\omega)}}{2a_{0}^{3(1+\omega)}\beta_{4}^{3h_{f}(1+\omega)}}-\frac{6Ch^{2}_{f}(-T)^{3\beta_{3}/2}}{\beta_{4}^{3/2}}-\frac{\mathcal{T}}{2}.\nonumber\\
% f(T,\mathcal{T})\simeq C_{6}T(-T)^{-\beta_{3}/2}+C_{7}(-T)^{\beta_{5}}+C_{8}(-T)^{3\beta_{3}/2}-\frac{1}{2}\mathcal{T},\nonumber\\
 f(T,\mathcal{T})\simeq -C_{6}(-T)^{1-\beta_{3}/2}+C_{7}(-T)^{\beta_{5}}+C_{8}(-T)^{3\beta_{3}/2}-\frac{1}{2}\mathcal{T},
\end{eqnarray}
where we have set 
\begin{eqnarray}
 C_{6}\equiv C\sqrt{\beta_{4}}, C_{7}\equiv\frac{\rho_{0}(1-5\omega)}{2a_{0}^{3(1+\omega)}\beta_{3}^{3h_{f}(1+\omega)}},C_{8}\equiv-\frac{6Ch^{2}_{f}}{\beta_{4}^{3/2}},\beta_{5}\equiv3\beta_{3}h_{f}(1+\omega).
\end{eqnarray}
The action (\ref{ord}) should describe a decelerated phase which can be interpreted as the one dominated by the non-relativistic ordinary matter.\\
\item For $h_{f}(1+\omega)>1/2\Rightarrow h_{f}>1/2(1+\omega)$, the action in this case can describe an accelerated universe or a decelerated universe depending on the value of the parameter $\omega$. Furthermore, one sees that the transition from the decelerated matter dominated phase to the accelerated dark energy dominated phase can occur . Then,  $\Phi$  behaves as 
 \begin{eqnarray}
 \Phi\sim3h_{f}\sqrt{-\frac{2}{T}},
\end{eqnarray}
leading to 
\begin{eqnarray}
\label{ord-som}
% f(T,\mathcal{T})\simeq-C(18h^{2}_{f})^{1/4}(-T)^{3/4}+\frac{\rho_{0}(1-5\omega)}{2a_{0}^{3(1+\omega)}(18h^{2}_{f})^{3h_{f}(1+\omega)/4}}(-T)^{3h_{f}(1+\omega)/2}-\frac{6Ch^{2}_{f}}{(18h^{2}_{f})^{3/8}}(-T)^{3/4}-\frac{\mathcal{T}}{2},\nonumber\\
%f(T,\mathcal{T})\simeq-C\left((18h^{2}_{f})^{1/4}+\frac{6h^{2}_{f}}{(18h^{2}_{f})^{3/8}}\right)(-T)^{3/4}+\frac{\rho_{0}(1-5\omega)}{2a_{0}^{3(1+\omega)}(18h^{2}_{f})^{3h_{f}(1+\omega)/4}}(-T)^{3h_{f}(1+\omega)/2}-\frac{\mathcal{T}}{2}.\\
  f(T,\mathcal{T})\simeq C_{4}(-T)^{3/4}+C_{5}(-T)^{3h_{f}(1+\omega)/2}-\frac{\mathcal{T}}{2},
\end{eqnarray}
with 
\begin{eqnarray}
 C_{4}\equiv-C\left((18h^{2}_{f})^{1/4}+\frac{6h^{2}_{f}}{(18h^{2}_{f})^{3/8}}\right), C_{5}\equiv\frac{\rho_{0}(1-5\omega)}{2a_{0}^{3(1+\omega)}(18h^{2}_{f})^{3h_{f}(1+\omega)/4}}.
\end{eqnarray}
\end{itemize}
%%%%%%%%%%%%%%%%%%%%%%%%%%%%%%%%%%%%%%%%%%%%%%%%%%%%%%%%%%%%%%%%%%%%%%%%%%%%%%%%%%%%%%%%%%%%%%%%%%%%%%%%%%%%%%%%%%%%%%%%%%%%%%%%%%%%%
%%%%%%%%%%%%%%%%%%%%%%%%%%%%%%%%%%%%%%%%%%%%%%%%%%%%%%%%%%%%%%%%%%%%%%%%%%%%%%%%%%%%%%%%%%%%%%%%%%%%%%%%%%%%%%%%%%%%%%%%%%%%%%%%%%%%%
%%%%%%%%%%%%%%%%%%%%%%%%%%%%%%%%%%%%%%%%%%%%%%%%%%%%%%%%%%%%%%%%%%%%%%%%%%%%%%%%%%%%%%%%%%%%%%%%%%%%%%%%%%%%%%%%%%%%%%%%%%%%%%%%%%%%%
%%%%%%%%%%%%%%%%%%%%%%%%%%%%%%%%%%%%%%%%%%%%%%%%%%%%%%%%%%%%%%%%%%%%%%%%%%%%%%%%%%%%%%%%%%%%%%%%%%%%%%%%%%%%%%%%%%%%%%%%%%%%%%%%%%%%%
%%%%%%%%%%%%%%%%%%%%%%%%%%%%%%%%%%%%%%%%%%%%%%%%%%%%%%%%%%%%%%%%%%%%%%%%%%%%%%%%%%%%%%%%%%%%%%%%%%%%%%%%%%%%%%%%%%%%%%%%%%%%%%%%%%%%%
%%%%%%%%%%%%%%%%%%%%%%%%%%%%%%%%%%%%%%%%%%%%%%%%%%%%%%%%%%%%%%%%%%%%%%%%%%%%%%%%%%%%%%%%%%%%%%%%%%%%%%%%%%%%%%%%%%%%%%%
%%%%%%%%%%%%%%%%%%%%%%%%%%%%%%%%%%%%%%%%%%%%%%%%%%%%%%%%%%%%%%%%%%%%%%%%%%%%%%%%%%%%%%%%%%%%%%%%%%%%%%%%%%%%%%%%%%%%%%%

\section{Second approach: more general case}
In order to generalise the previous results we will consider that the second side of the equation  (\ref{12}) does not vanish, setting it to a constant  $\lambda$, the so-called separation of variables, i,e, 
\begin{eqnarray}
\label{G1}
8H\dot{P}_{1}(\Phi)+4\dot{H}P_{1}(\Phi)&=&\lambda,\\
\label{G2}
P_{2}(\Phi)=-\frac{1}{2}\left(\frac{\lambda}{\rho(1+\omega)}+1\right)&=&-\frac{1}{2}\left(\frac{\lambda a^{3(1+\omega)}}{\rho_{0}(1+\omega)}+1\right).
\end{eqnarray}
From the relation (\ref{20}), we obtain,
\begin{eqnarray}
 P_{1}(\Phi)=\frac{1}{g_{0}^{1/2}}\left(1+\frac{g_{1}}{g_{0}\Phi}\right)^{-1/2}\left\{C+\frac{\lambda}{8}\left[\frac{\Phi}{g_{0}^{1/2}}\left(1+\frac{g_{1}}{g_{0}\Phi}\right)^{1/2}\right.\right.\nonumber\\
 \left.\left.-\frac{g_{1}}{g_{0}^{3/2}}\ln\left(g_{0}\Phi^{1/2}\left[1+\left(1+\frac{g_{1}}{g_{0}\Phi}\right)^{1/2}\right]\right)\right]\right\},\\
 P_{2}(\Phi)=-\frac{1}{2}\left(1+\frac{\lambda a_{0}^{3(1+\omega)}e^{3g_{0}(1+\omega)\Phi}\Phi^{3g_{1}(1+\omega)}}{\rho_{0}(1+\omega)}\right)
\end{eqnarray}
By inserting these two expressions into the relation (\ref{10}), it appears 
\begin{eqnarray}
 Q(\Phi)=-\left(\frac{1+5\omega}{1+\omega}\right)\frac{\lambda}{2}+\frac{\rho_{0}(1-5\omega)}{2a_{0}^{3(1+\omega)}}e^{-3g_{0}(1+\omega)\Phi}\Phi^{-3g_{1}(1+\omega)}\nonumber\\
 -6g_{0}^{3/2}\left(1+\frac{g_{1}}{g_{0}\Phi}\right)^{3/2}\left\{C+\frac{\lambda}{8}\left[\frac{\Phi}{g_{0}^{1/2}}\left(1+\frac{g_{1}}{g_{0}\Phi}\right)^{1/2}\right.\right.\nonumber\\
 \left.\left.-\frac{g_{1}}{g_{0}^{3/2}}\ln\left(g_{0}\Phi^{1/2}\left[1+\left(1+\frac{g_{1}}{g_{0}\Phi}\right)^{1/2}\right]\right)\right]\right\}
\end{eqnarray}
\begin{enumerate}
 \item When $0<\Phi\ll1$, the relation (\ref{b}) becomes
 \begin{eqnarray}
  -\frac{3\mathcal{T}\lambda a_{0}^{3(1+\omega)g_{0}}}{2\rho_{0}}\Phi^{3g_{1}(1+\omega)}-\frac{3\rho_{0}(1-5\omega)(1+\omega)}{2a_{0}^{3(1+\omega)}}\Phi^{-3(1+\omega)}\nonumber\\
  -\frac{3\rho_{0}(1-5\omega)(1+\omega)}{2a_{0}^{3(1+\omega)}}\Phi^{-3(1+\omega)-1}+\left(C-\frac{\lambda g_{1}\ln(g_{0}g_{1})}{16g_{0}^{3/2}}\right)\frac{T}{2g_{1}^{1/2}}\Phi^{-1/2}\nonumber\\
  -\frac{3g_{1}^{2}(1-g_{0}^{1/2})\lambda}{8g_{0}}\Phi^{-2}+2g_{1}^{3/2}\left(C-\frac{\lambda g_{1}\ln(g_{0}g_{1})}{16g_{0}^{3/2}}\right)\Phi^{-5/2}+\frac{\lambda(1-g_{0}^{1/2})T}{16g_{0}}\simeq0
 \end{eqnarray}
 By simplification, we perform an expansion to the first order for establishing the equation. Thus 
\begin{itemize}
 \item If $3g_{1}(1+\omega)+1>5/2\Rightarrow g_{1}(1+\omega)>1/2$, then 
 \begin{eqnarray}
 \Phi\sim\alpha_{1}\left(-\frac{1}{T}\right)^{\alpha_{0}},
\end{eqnarray}
with 
\begin{eqnarray}
 \alpha_{0}\equiv\frac{1}{3g_{1}(1+\omega)},
 \alpha_{1}\equiv\left(\frac{3\rho_{0}(1-5\omega)(1+\omega)}{\lambda a_{0}^{3(1+\omega)(\sqrt{g_{0}}-1)}}\right)^{\alpha_{0}}.
\end{eqnarray}
Therefore 
%\begin{eqnarray}
% f(T,\mathcal{T})\simeq\left\{\frac{CT}{g_{0}^{1/2}}\left(1+\frac{g_{1}(-T)^{\alpha_{0}}}{g_{0}\alpha_{1}}\right)^{-1/2}-6g_{0}^{3/2}\left(1+\frac{g_{1}(-T)^{\alpha_{0}}}{g_{0}\alpha_{1}}\right)^{3/2}\right\}\nonumber\\
% \times\left\{C+\frac{\lambda}{8}\left[\frac{\alpha_{1}(-T)^{-\alpha_{0}}}{g_{0}^{1/2}}\left(1+\frac{g_{1}(-T)^{\alpha_{0}}}{g_{0}\alpha_{1}}\right)^{1/2}\right.\right.\nonumber\\
% \left.\left.-\frac{g_{1}}{g_{0}^{3/2}}\ln\left(g_{0}\alpha_{1}^{1/2}(-T)^{-\alpha_{0}/2}\left[1+\left(1+\frac{g_{1}(-T)^{\alpha_{0}}}{g_{0}\alpha_{1}}\right)^{1/2}\right]\right)\right]\right\}\nonumber\\
%-\left(\frac{1+5\omega}{1+\omega}\right)\frac{\lambda}{2}+\frac{\rho_{0}(1-5\omega)(-T)^{3g_{1}\alpha_{0}(1+\omega)}}{2a_{0}^{3(1+\omega)}\alpha_{1}^{3g_{1}(1+\omega)}}Exp\left(-3g_{0}\alpha_{1}(1+\omega)(-T)^{\alpha_{0}}\right)\nonumber\\
%-\frac{\mathcal{T}}{2}-\frac{\lambda a_{0}^{3(1+\omega)}\mathcal{T}Exp\left(3g_{0}\alpha_{1}(1+\omega)(-T)^{-\alpha_{0}}\right)\alpha_{1}^{3g_{1}(1+\omega)}}{2\rho_{0}(1+\omega)(-T)^{3\alpha_{0}g_{1}(1+\omega)}}.
%\end{eqnarray}
\begin{eqnarray}
 f(T,\mathcal{T})\simeq\Psi_{2}(T)+\Psi_{3}(T)+\Psi_{4}(T,\mathcal{T})-\frac{\mathcal{T}}{2}-\left(\frac{1+5\omega}{1+\omega}\right)\frac{\lambda}{2},
\end{eqnarray}
with 
\begin{eqnarray}
 \Psi_{2}(T)\equiv\left\{\frac{T}{g_{0}^{1/2}}\left(1+\frac{g_{1}(-T)^{\alpha_{0}}}{g_{0}\alpha_{1}}\right)^{-1/2}-6g_{0}^{3/2}\left(1+\frac{g_{1}(-T)^{\alpha_{0}}}{g_{0}\alpha_{1}}\right)^{3/2}\right\}\nonumber\\
 \times\left\{C+\frac{\lambda}{8}\left[\frac{\alpha_{1}(-T)^{-\alpha_{0}}}{g_{0}^{1/2}}\left(1+\frac{g_{1}(-T)^{\alpha_{0}}}{g_{0}\alpha_{1}}\right)^{1/2}\right.\right.\nonumber\\
 \left.\left.-\frac{g_{1}}{g_{0}^{3/2}}\ln\left(g_{0}\alpha_{1}^{1/2}(-T)^{-\alpha_{0}/2}\left[1+\left(1+\frac{g_{1}(-T)^{\alpha_{0}}}{g_{0}\alpha_{1}}\right)^{1/2}\right]\right)\right]\right\},\nonumber\\
 \Psi_{3}(T)\equiv\frac{\rho_{0}(1-5\omega)(-T)^{3g_{1}\alpha_{0}(1+\omega)}e^{-3g_{0}\alpha_{1}(1+\omega)(-T)^{\alpha_{0}}}}{2a_{0}^{3(1+\omega)}\alpha_{1}^{3g_{1}(1+\omega)}}\\
 \Psi_{4}(T,\mathcal{T})\equiv-\frac{\lambda a_{0}^{3(1+\omega)}\mathcal{T}e^{3g_{0}\alpha_{1}(1+\omega)(-T)^{-\alpha_{0}}}\alpha_{1}^{3g_{1}(1+\omega)}}{2\rho_{0}(1+\omega)(-T)^{3\alpha_{0}g_{1}(1+\omega)}}.\nonumber 
\end{eqnarray}
\item If  $3g_{1}(1+\omega)+1<5/2\Rightarrow g_{1}(1+\omega)<1/2$, then
\begin{eqnarray}
 \Phi\sim\alpha_{2}\left(-\frac{1}{T}\right)^{2/5},\\
 \alpha_{2}\equiv\left(\frac{2g_{1}^{3/2}(16g_{0}^{3/2}-g_{1}\lambda\ln(g_{0}g_{1}))}{g_{0}^{1/2}(g_{0}^{1/2}\lambda-1)}\right)^{2/5}.
\end{eqnarray}
Therefore
%\begin{eqnarray}
% f(T,\mathcal{T})\simeq\left\{\frac{CT}{g_{0}^{1/2}}\left(1+\frac{g_{1}(-T)^{2/5}}{g_{0}\alpha_{2}}\right)^{-1/2}-6g_{0}^{3/2}\left(1+\frac{g_{1}(-T)^{2/5}}{g_{0}\alpha_{2}}\right)^{3/2}\right\}\nonumber\\
% \times\left\{C+\frac{\lambda}{8}\left[\frac{\alpha_{2}}{g_{0}^{1/2}(-T)^{2/5}}\left(1+\frac{g_{1}(-T)^{2/5}}{g_{0}\alpha_{2}}\right)^{1/2}\right.\right.\nonumber\\
% \left.\left.-\frac{g_{1}}{g_{0}^{3/2}}\ln\left(g_{0}\sqrt{\alpha_{2}}(-T)^{-1/5}\left[1+\left(1+\frac{g_{1}(-T)^{2/5}}{g_{0}\alpha_{2}}\right)^{1/2}\right]\right)\right]\right\}\nonumber\\
% -\left(\frac{1+5\omega}{1+\omega}\right)\frac{\lambda}{2}+\frac{\rho_{0}(1-5\omega)(-T)^{6g_{1}(1+\omega)/5}}{2a_{0}^{3(1+\omega)}\alpha_{2}^{3g_{1}(1+\omega)}}Exp\left(-3g_{0}\alpha_{2}(1+\omega)(-T)^{-2/5}\right)\nonumber\\
% -\frac{\mathcal{T}}{2}-\frac{\lambda\mathcal{T}a_{0}^{3(1+\omega)}\alpha_{2}^{3g_{1}(1+\omega)}Exp\left(3g_{0}\alpha_{2}(1+\omega)(-T)^{-2/5}\right)}{2\rho_{0}(1+\omega)(-T)^{6g_{1}(1+\omega)/5}}.
%\end{eqnarray}
\begin{eqnarray}
 f(T,\mathcal{T})\simeq\Psi_{5}(T)+\Psi_{6}(T)+\Psi_{7}(T,\mathcal{T})-\frac{\mathcal{T}}{2}-\left(\frac{1+5\omega}{1+\omega}\right)\frac{\lambda}{2},
\end{eqnarray}
with 
\begin{eqnarray}
 \Psi_{5}(T)\equiv\left\{\frac{T}{g_{0}^{1/2}}\left(1+\frac{g_{1}(-T)^{2/5}}{g_{0}\alpha_{2}}\right)^{-1/2}-6g_{0}^{3/2}\left(1+\frac{g_{1}(-T)^{2/5}}{g_{0}\alpha_{2}}\right)^{3/2}\right\}\nonumber\\
 \times\left\{C+\frac{\lambda}{8}\left[\frac{\alpha_{2}}{g_{0}^{1/2}(-T)^{2/5}}\left(1+\frac{g_{1}(-T)^{2/5}}{g_{0}\alpha_{2}}\right)^{1/2}\right.\right.\nonumber\\
 \left.\left.-\frac{g_{1}}{g_{0}^{3/2}}\ln\left(g_{0}\sqrt{\alpha_{2}}(-T)^{-1/5}\left[1+\left(1+\frac{g_{1}(-T)^{2/5}}{g_{0}\alpha_{2}}\right)^{1/2}\right]\right)\right]\right\}\nonumber\\
 \Psi_{6}(T)\equiv\frac{\rho_{0}(1-5\omega)(-T)^{6g_{1}(1+\omega)/5}e^{-3g_{0}\alpha_{2}(1+\omega)(-T)^{-2/5}}}{2a_{0}^{3(1+\omega)}\alpha_{2}^{3g_{1}(1+\omega)}},\\
\Psi_{7}(T,\mathcal{T})\equiv-\frac{\lambda\mathcal{T}a_{0}^{3(1+\omega)}\alpha_{2}^{3g_{1}(1+\omega)}e^{3g_{0}\alpha_{2}(1+\omega)(-T)^{-2/5}}}{2\rho_{0}(1+\omega)(-T)^{6g_{1}(1+\omega)/5}}.
\end{eqnarray}
\end{itemize}
\item $\Phi\longrightarrow+\infty$, in this case one obtains 
\begin{eqnarray}
 Q(\Phi)\simeq-\left(\frac{1+5\omega}{1+\omega}\right)\frac{\lambda}{2}-6g_{0}^{3/2}\left(1+\frac{g_{1}}{g_{0}\Phi}\right)^{3/2}\left\{C+\frac{\lambda}{8}\left[\frac{\Phi}{g_{0}^{1/2}}\left(1+\frac{g_{1}}{g_{0}\Phi}\right)^{1/2}\right.\right.\nonumber\\
 \left.\left.-\frac{g_{1}}{g_{0}^{3/2}}\ln\left(g_{0}\Phi^{1/2}\left[1+\left(1+\frac{g_{1}}{g_{0}\Phi}\right)^{1/2}\right]\right)\right]\right\},
\end{eqnarray}
such that 
\begin{eqnarray}
 -\frac{3\mathcal{T}\lambda a_{0}^{3(1+\omega)}g_{0}}{2\rho_{0}}\left(1+\frac{g_{1}}{g_{0}\Phi}\right)e^{3g_{0}(1+\omega)\Phi}\Phi^{3g_{1}(1+\omega)+3}+\frac{g_{1}TC\Phi}{16g_{0}^{3/2}}+9Cg_{1}g_{0}^{1/2}\Phi\nonumber\\
 +\frac{g_{1}^{2}\lambda(1-2\ln2g_{0})}{32g_{0}^{3}}(T+18g_{0}^{2})\Phi-\Phi\left(\frac{g_{1}T\lambda}{16g_{0}^{3/2}}+\frac{9g_{1}g_{0}^{1/2}\lambda}{8}\right)\left(\frac{\Phi}{g_{0}^{1/2}}+\frac{g_{1\ln(\Phi)}}{2g_{0}^{3/2}}\right)\nonumber\\
 -\left(\frac{Tg_{1}\lambda}{3g_{0}^{3/2}}+\frac{9g_{1}g_{0}^{1/2}\lambda}{16}\right)\frac{2\Phi^{2}}{g_{0}^{1/2}}+\frac{\lambda\Phi}{8g_{0}^{1/2}}\left(9g_{1}^{2}g_{0}^{1/2}-\frac{3g_{1}^{2}T}{4g_{0}^{5/2}}\right)+\left(\frac{Tg_{1}\lambda}{3g_{0}^{3/2}}+\frac{9g_{1}g_{0}^{1/2}\lambda}{16}\right)\frac{2\Phi}{g_{0}^{1/2}}\nonumber\\
 -\frac{g_{1}\lambda}{16g_{0}^{3/2}}\left(9g_{1}^{2}g_{0}^{1/2}-\frac{3g_{1}^{2}T}{4g_{0}^{5/2}}\right)\ln\Phi+\left(\frac{T\lambda}{16g_{0}^{1/2}}-\frac{3g_{0}^{3/2}\lambda}{8}\right)\left(\frac{2\Phi^{3}}{g_{0}^{1/2}}-\frac{g_{1}\Phi^{2}}{g_{0}^{3/2}}\right)-\frac{3g_{1}^{2}\lambda\Phi}{8g_{0}}\nonumber\\
 +\left(\frac{g_{1}\lambda(1-2\ln2g_{0})}{16g_{0}^{3/2}}+C\right)\left(9g_{1}^{2}g_{0}^{1/2}-\frac{3g_{1}^{2}T}{4g_{0}^{5/2}}\right)+\frac{3g_{1}\lambda}{16g_{0}^{2}}\sim0
\end{eqnarray}
It appears that  
\begin{eqnarray}
% \frac{3\mathcal{T}\lambda a_{0}^{3(1+\omega)}g_{0}}{2\rho_{0}}e^{3g_{0}(1+\omega)\Phi}\Phi^{3g_{1}(1+\omega)+3}\sim\left(\frac{g_{1}\lambda(1-2\ln2g_{0})}{16g_{0}^{3/2}}+C\right)\left(9g_{1}^{2}g_{0}^{1/2}-\frac{3g_{1}^{2}T}{4g_{0}^{5/2}}\right)+\frac{3g_{1}\lambda}{16g_{0}^{2}}
% \frac{3\mathcal{T}\lambda a_{0}^{3(1+\omega)}g_{0}}{2\rho_{0}}e^{3g_{0}(1+\omega)\Phi}\sim\left(\frac{g_{1}\lambda(1-2\ln2g_{0})}{16g_{0}^{3/2}}+C\right)\left(9g_{1}^{2}g_{0}^{1/2}-\frac{3g_{1}^{2}T}{4g_{0}^{5/2}}\right)+\frac{3g_{1}\lambda}{16g_{0}^{2}},\nonumber\\
 e^{3g_{0}(1+\omega)\Phi}\sim\frac{2\rho_{0}}{3\mathcal{T}\lambda a_{0}^{3(1+\omega)}g_{0}}\left(\left(\frac{g_{1}\lambda(1-2\ln2g_{0})}{16g_{0}^{3/2}}+C\right)\left(9g_{1}^{2}g_{0}^{1/2}-\frac{3g_{1}^{2}T}{4g_{0}^{5/2}}\right)+\frac{3g_{1}\lambda}{16g_{0}^{2}}\right)
\end{eqnarray}
\begin{eqnarray}
 \Phi\sim\alpha_{3}\ln\left(\alpha_{4}\left(-\frac{T}{\mathcal{T}}\right)+\frac{\alpha_{5}}{\mathcal{T}}\right),
\end{eqnarray}
with 
\begin{eqnarray}
 \alpha_{3}\equiv\frac{1}{3g_{0}(1+\omega)},\alpha_{4}\equiv\frac{\rho_{0}g_{1}^{2}}{\lambda a_{0}^{3(1+\omega)}g_{0}^{7/2}}\left(\frac{g_{1}\lambda(1-2\ln2g_{0})}{16g_{0}^{3/2}}+C\right)\\
 \alpha_{5}\equiv\frac{2\rho_{0}}{3\lambda a_{0}^{3(1+\omega)}g_{0}}\left(9g_{1}^{2}g_{0}^{1/2}\left(\frac{g_{1}\lambda(1-2\ln2g_{0})}{16g_{0}^{3/2}}+C\right)+\frac{3g_{1}\lambda}{16g_{0}^{2}}\right).
\end{eqnarray}
Therefore 
%\begin{eqnarray}
% f(T,\mathcal{T})\simeq\left\{\frac{T}{g_{0}^{1/2}}\left(1+\frac{g_{1}}{g_{0}\alpha_{3}\ln\left(\alpha_{4}\left(-\frac{T}{\mathcal{T}}\right)+\frac{\alpha_{5}}{\mathcal{T}}\right)}\right)^{-1/2}\right.\nonumber\\
% \left.-6g_{0}^{3/2}\left(1+\frac{g_{1}}{g_{0}\alpha_{3}\ln\left(\alpha_{4}\left(-\frac{T}{\mathcal{T}}\right)+\frac{\alpha_{5}}{\mathcal{T}}\right)}\right)^{3/2}\right\}\nonumber\\
% \times\left\{C+\frac{\lambda}{8}\left[\frac{\alpha_{3}\ln\left(\alpha_{4}\left(-\frac{T}{\mathcal{T}}\right)+\frac{\alpha_{5}}{\mathcal{T}}\right)}{g_{0}^{1/2}}\left(1+\frac{g_{1}}{g_{0}\alpha_{3}\ln\left(\alpha_{4}\left(-\frac{T}{\mathcal{T}}\right)+\frac{\alpha_{5}}{\mathcal{T}}\right)}\right)^{1/2}\right.\right.\nonumber\\
% \left.\left.-\frac{g_{1}}{g_{0}^{3/2}}\ln\left(\frac{g_{0}\sqrt{\alpha_{3}}}{2} \ln\left(\alpha_{4}\left(-\frac{T}{\mathcal{T}}\right)+\frac{\alpha_{5}}{\mathcal{T}}\right)\left[1+\left(1+\frac{g_{1}}{g_{0}\alpha_{3}\ln\left(\alpha_{4}\left(-\frac{T}{\mathcal{T}}\right)+\frac{\alpha_{5}}{\mathcal{T}}\right)}\right)^{1/2}\right]\right)\right]\right\}\nonumber\\
% -\left(\frac{1+5\omega}{1+\omega}\right)\frac{\lambda}{2}-\frac{\mathcal{T}}{2}-\frac{\lambda\mathcal{T} a_{0}^{3(1+\omega)}\left(\alpha_{4}\left(-\frac{T}{\mathcal{T}}\right)+\frac{\alpha_{5}}{\mathcal{T}}\right)^{3g_{0}\alpha_{3}(1+\omega)}\alpha_{3}^{3g_{1}(1+\omega)}3g_{1}\left(\ln\left(\alpha_{4}\left(-\frac{T}{\mathcal{T}}\right)+\frac{\alpha_{5}}{\mathcal{T}}\right)\right)^{3g_{1}(1+\omega)}}{2\rho_{0}}
%\end{eqnarray}
\begin{eqnarray}
 f(T,\mathcal{T})\simeq\Psi_{8}(T,\mathcal{T})+\Psi_{9}(T,\mathcal{T})-\frac{\mathcal{T}}{2}-\left(\frac{1+5\omega}{1+\omega}\right)\frac{\lambda}{2},
\end{eqnarray}
with 
\begin{eqnarray}
 \Psi_{8}(T,\mathcal{T})\equiv\left\{\frac{T}{g_{0}^{1/2}}\left(1+\frac{g_{1}}{g_{0}\alpha_{3}\ln\left(\alpha_{4}\left(-\frac{T}{\mathcal{T}}\right)+\frac{\alpha_{5}}{\mathcal{T}}\right)}\right)^{-1/2}\right.\nonumber\\
 \left.-6g_{0}^{3/2}\left(1+\frac{g_{1}}{g_{0}\alpha_{3}\ln\left(\alpha_{4}\left(-\frac{T}{\mathcal{T}}\right)+\frac{\alpha_{5}}{\mathcal{T}}\right)}\right)^{3/2}\right\}\nonumber\\
 \times\left\{C+\frac{\lambda}{8}\left[\frac{\alpha_{3}\ln\left(\alpha_{4}\left(-\frac{T}{\mathcal{T}}\right)+\frac{\alpha_{5}}{\mathcal{T}}\right)}{g_{0}^{1/2}}\left(1+\frac{g_{1}}{g_{0}\alpha_{3}\ln\left(\alpha_{4}\left(-\frac{T}{\mathcal{T}}\right)+\frac{\alpha_{5}}{\mathcal{T}}\right)}\right)^{1/2}\right.\right.\nonumber\\
 \left.\left.-\frac{g_{1}}{g_{0}^{3/2}}\ln\left(\frac{g_{0}\sqrt{\alpha_{3}}}{2} \ln\left(\alpha_{4}\left(-\frac{T}{\mathcal{T}}\right)+\frac{\alpha_{5}}{\mathcal{T}}\right)\left[1+\left(1+\frac{g_{1}}{g_{0}\alpha_{3}\ln\left(\alpha_{4}\left(-\frac{T}{\mathcal{T}}\right)+\frac{\alpha_{5}}{\mathcal{T}}\right)}\right)^{1/2}\right]\right)\right]\right\},\\
 \Psi_{9}(T,\mathcal{T})\equiv-\frac{3\lambda\alpha_{3}^{3g_{1}(1+\omega)}a_{0}^{3(1+\omega)}g_{1}}{2\rho_{0}}\nonumber\\
 \times\left[\mathcal{T}\left(\alpha_{4}\left(-\frac{T}{\mathcal{T}}\right)+\frac{\alpha_{5}}{\mathcal{T}}\right)^{3g_{0}\alpha_{3}(1+\omega)}\left(\ln\left(\alpha_{4}\left(-\frac{T}{\mathcal{T}}\right)+\frac{\alpha_{5}}{\mathcal{T}}\right)\right)^{3g_{1}(1+\omega)}\right].
\end{eqnarray}

\end{enumerate}
%%%%%%%%%%%%%%%%%%%%%%%%%%%%%%%%%%%%%%%%%%%%%%%%%%%%%%%%%%%%%%%%%%%%%%%%%%%%%%%%%%%%%%%%%%%%%%%%%%%%%%%%%%%%%%%%%%%%%%%%%%%%
%%%%%%%%%%%%%%%%%%%%%%%%%%%%%%%%%%%%%%%%%%%%%%%%%%%%%%%%%%%%%%%%%%%%%%%%%%%%%%%%%%%%%%%%%%%%%%%%%%%%%%%%%%%%%%%%%%%%%%%%%%%%
%%%%%%%%%%%%%%%%%%%%%%%%%%%%%%%%%%%%%%%%%%%%%%%%%%%%%%%%%%%%%%%%%%%%%%%%%%%%%%%%%%%%%%%%%%%%%%%%%%%%%%%%%%%%%%%%%%%%%%%%%%%%
%%%%%%%%%%%%%%%%%%%%%%%%%%%%%%%%%%%%%%%%%%%%%%%%%%%%%%%%%%%%%%%%%%%%%%%%%%%%%%%%%%%%%%%%%%%%%%%%%%%%%%%%%%%%%%%%%%%%%%%%%%%%
%%%%%%%%%%%%%%%%%%%%%%%%%%%%%%%%%%%%%%%%%%%%%%%%%%%%%%%%%%%%%%%%%%%%%%%%%%%%%%%%%%%%%%%%%%%%%%%%%%%%%%%%%%%%%%%%%%%%%%%%%%%%
Now we will consider the definition (\ref{24}). This new approach allows to obtain from  (\ref{G1}) and  (\ref{G2}), the following  expressions.
\begin{eqnarray}
 P_{1}(\Phi)&=&P_{0}\Phi^{1/2}+\frac{2\lambda\Phi^{2}}{3h_{f}}+k_{0}\Phi^{1/2}\Bigg[Arctan\left(1-k_{1}\sqrt{\Phi}\right)-Arctan\left(1+k_{1}\sqrt{\Phi}\right)\Bigg]\nonumber\\
 &&-\frac{k_{0}\Phi^{1/2}}{2}\Bigg[\ln\left(\sqrt{h_{i}}-(qh_{i}h_{f})^{1/4}\sqrt{2\Phi}+\sqrt{qh_{f}}\Phi\right)\Bigg.\nonumber\\
&& \Bigg.-\ln\left(\sqrt{h_{i}}+(qh_{i}h_{f})^{1/4}\sqrt{2\Phi}+\sqrt{qh_{f}}\Phi\right)\Bigg], \\
P_{2}&=&-\frac{1}{2}\left(\frac{\lambda a_{0}^{3(1+\omega)}\Phi^{3h(\Phi)(1+\omega)}}{\rho_{0}(1+\omega)}+1\right).
\end{eqnarray}
with
\begin{eqnarray}
 k_{0}\equiv\frac{\lambda\sqrt{2}(h_{i}-h_{f})}{2q^{3/4}h_{f}^{7/4}h_{i}^{1/4}}, k_{1}\equiv\left(\frac{4qh_{f}}{h_{i}}\right)^{1/4}.
\end{eqnarray}
Therefore, from the relation (\ref{10}), it appears that 
\begin{eqnarray}
 Q(\Phi)=\frac{\rho_{0}(1-5\omega)}{2a_{0}^{3(1+\omega)}}\Phi^{-3h(\Phi)(1+\omega)}-6P_{0}h^{2}(\Phi)\Phi^{-3/2}\nonumber\\
 -6P_{0}h^{2}(\Phi)k_{0}\Phi^{-3/2}\Bigg[Arctan(1-k_{1}\sqrt{\Phi})-Arctan(1+k_{1}\sqrt{\Phi})\Bigg]\nonumber\\
 +3k_{0}h^{2}(\Phi)\Phi^{-3/2}\Bigg[\ln\left(\sqrt{h_{i}}-(4qh_{i}h_{f})^{1/4}\sqrt{\Phi}+\sqrt{qh_{f}}\Phi\right)\Bigg.\nonumber\\
\Bigg. -\ln\left(\sqrt{h_{i}}+(4qh_{i}h_{f})^{1/4}\sqrt{\Phi}+\sqrt{qh_{f}}\Phi\right)\Bigg].
\end{eqnarray}
\begin{enumerate}
 \item When $\Phi\longrightarrow0$, one gets 
 \begin{eqnarray}
 P_{2}(\Phi)\simeq-\frac{1}{2}\left(1+\frac{\lambda a_{0}^{3(1+\omega)}}{\rho_{0}(1+\omega)}\Phi^{3h_{i}(1+\omega)}\right),
\end{eqnarray}
and
\begin{eqnarray}
  Q(\Phi)\simeq\frac{\rho_{0}(1-5\omega)}{2a_{0}^{3(1+\omega)}}\Phi^{-3h_{i}(1+\omega)}-6P_{0}h^{2}_{i}\Phi^{-3/2}\nonumber\\
 -6P_{0}h^{2}_{i}k_{0}\Phi^{-3/2}\Bigg[Arctan(1-k_{1}\sqrt{\Phi})-Arctan(1+k_{1}\sqrt{\Phi})\Bigg]\nonumber\\
 +3k_{0}h^{2}_{i}\Phi^{-3/2}\Bigg[\ln\left(\sqrt{h_{i}}-(4qh_{i}h_{f})^{1/4}\sqrt{\Phi}+\sqrt{qh_{f}}\Phi\right)\Bigg.\nonumber\\
\Bigg. -\ln\left(\sqrt{h_{i}}+(4qh_{i}h_{f})^{1/4}\sqrt{\Phi}+\sqrt{qh_{f}}\Phi\right)\Bigg].
\end{eqnarray}
The relation (\ref{b}) becomes
\begin{eqnarray}
\frac{4T\lambda}{3h_{f}}-\frac{3h_{i}\rho_{0}(1+\omega)(1-5\omega)}{2a_{0}^{3(1+\omega)}}\Phi^{-3h_{i}(1+\omega)-2}-\frac{3h_{i}a_{0}^{3(1+\omega)}}{2\rho_{0}}\Phi^{3(1+\omega)-2}\nonumber\\
+\left(-k_{0}k_{1}T-\frac{k_{0}T}{2}+\frac{k_{0}T(4qh_{i}h_{f})^{1/4}}{2\sqrt{h_{i}}}+\frac{k_{0}T\sqrt{qh_{f}}(4qh_{i}h_{f})^{1/4}}{2}\right)\Phi^{-1}\nonumber\\
+\frac{P_{0}T}{2}\Phi^{-3/2}+\left(3k_{0}h_{i}^{2}\sqrt{qh_{f}}(4qh_{i}h_{f})^{1/4}+3k_{1}^{2}k_{0}h_{i}^{2}\right)\Phi^{-2}\nonumber\\
-\left(12k^{2}_{1}k_{0}+9k_{0}h_{i}^{3/2}(4qh_{i}h_{f})^{1/4}-3k_{0}h_{i}^{5/2}(4qh_{i}h_{f})^{1/4}\right)\Phi^{-3}\sim0
\end{eqnarray}
In order to establish this equation, we have performed the expansion to the first order of the non-linear quantities by simplification.
\begin{itemize}
\item If  $3h_{i}(1+\omega)+2>3\Rightarrow h_{i}>\frac{1}{3(1+\omega)}$, then
\begin{eqnarray}
% -\frac{3h_{i}\rho_{0}(1+\omega)(1-5\omega)}{2a_{0}^{3(1+\omega)}}\Phi^{-3h_{i}(1+\omega)-2}\sim-\frac{4T\lambda}{3h_{f}},
\Phi&\sim&\gamma_{1}\left(-\frac{1}{T}\right)^{\gamma_{0}},\\
\gamma_{0}&\equiv&\frac{1}{3h_{i}(1+\omega)},\nonumber\\
\gamma_{1}&\equiv&\left(\frac{9h_{i}h_{f}(1+\omega)(1-5\omega)}{8a_{0}^{3(1+\omega)\lambda}}\right)^{\gamma_{0}}.\nonumber
\end{eqnarray}
Thus 
%\begin{eqnarray}
% f(T,\mathcal{T})&\simeq& -C\sqrt{\gamma_{1}}(-T)^{1-\gamma_{0}/2}+\frac{2\lambda\gamma_{1}^{2}T^{-2\gamma_{0}+1}}{3h_{f}}\nonumber\\
% &&-\left(6Ch^{2}_{i}k_{0}\gamma_{1}^{-3/2}(-T)^{3\gamma_{0}/2}+k_{0}\sqrt{\gamma_{1}}(-T)^{1-\gamma_{0}/2}\right)\times\nonumber\\
% &&\left(Arctan\left(1-k_{1}\sqrt{\gamma_{1}}(-T)^{-\gamma_{0}/2}\right)-Arctan\left(1+k_{1}\sqrt{\gamma_{1}}(-T)^{-\gamma_{0}/2}\right)\right)\nonumber\\
%&&+\left(3k_{0}h^{2}_{i}\gamma_{1}^{-3/2}(-T)^{3\gamma_{0}/2}\Phi^{-3/2}+\frac{k_{0}\sqrt{\gamma_{1}}(-T)^{1-\gamma_{0}/2}}{2}\right)\times\nonumber\\
%&&\left(\ln\left(\sqrt{h_{i}}-(4qh_{i}h_{f})^{1/4}\sqrt{\gamma_{1}}(-T)^{-\gamma_{0}/2}+\sqrt{qh_{f}}\gamma_{1}(-T)^{-\gamma_{0}}\right)\right.\nonumber\\
%&& \left.-\ln\left(\sqrt{h_{i}}+(4qh_{i}h_{f})^{1/4}\sqrt{\gamma_{1}}(-T)^{-\gamma_{0}/2}+\sqrt{qh_{f}}\gamma_{1}(-T)^{-\gamma_{0}}\right)\right)\nonumber\\
%&&+\frac{\rho_{0}(1-5\omega)}{2a_{0}^{3(1+\omega)}\gamma_{1}^{3h_{i}(1+\omega)}}(-T)^{3h_{i}\gamma_{0}(1+\omega)}-6Ch^{2}_{i}\gamma_{1}^{-3/2}(-T)^{3\gamma_{0}/2}\nonumber\\
%&&-\frac{\mathcal{T}\lambda a_{0}^{3(1+\omega)}\gamma_{1}^{3h_{i}(1+\omega)}(-T)^{-3h_{i}\gamma_{0}(1+\omega)}}{2\rho_{0}(1+\omega)}-\frac{1}{2}\mathcal{T}.
%\end{eqnarray}
\begin{eqnarray}
 f(T,\mathcal{T})\simeq\zeta_{0}(T)+\zeta_{1}(T)+\zeta_{2}(T,\mathcal{T})-\frac{1}{2}\mathcal{T}.
\end{eqnarray}
where 
\begin{eqnarray}
 \zeta_{0}(T)\equiv-P_{0}\sqrt{\gamma_{1}}(-T)^{1-\gamma_{0}/2}+\frac{2\lambda\gamma_{1}^{2}T^{-2\gamma_{0}+1}}{3h_{f}}+\frac{\rho_{0}(1-5\omega)}{2a_{0}^{3(1+\omega)}\gamma_{1}^{3h_{i}(1+\omega)}}(-T)^{3h_{i}\gamma_{0}(1+\omega)}\nonumber\\
 -6P_{0}h^{2}_{i}\gamma_{1}^{-3/2}(-T)^{3\gamma_{0}/2},\\
 \zeta_{1}(T)\equiv-\Bigg[6P_{0}h^{2}_{i}k_{0}\gamma_{1}^{-3/2}(-T)^{3\gamma_{0}/2}+k_{0}\sqrt{\gamma_{1}}(-T)^{1-\gamma_{0}/2}\Bigg]\times\nonumber\\
 \Bigg[Arctan\left(1-k_{1}\sqrt{\gamma_{1}}(-T)^{-\gamma_{0}/2}\right)-Arctan\left(1+k_{1}\sqrt{\gamma_{1}}(-T)^{-\gamma_{0}/2}\right)\Bigg]\nonumber\\
+\Bigg[3k_{0}h^{2}_{i}\gamma_{1}^{-3/2}(-T)^{3\gamma_{0}/2}\Phi^{-3/2}+\frac{k_{0}\sqrt{\gamma_{1}}(-T)^{1-\gamma_{0}/2}}{2}\Bigg]\times\nonumber\\
\Bigg[\ln\left(\sqrt{h_{i}}-(4qh_{i}h_{f})^{1/4}\sqrt{\gamma_{1}}(-T)^{-\gamma_{0}/2}+\sqrt{qh_{f}}\gamma_{1}(-T)^{-\gamma_{0}}\right)\Bigg.\nonumber\\
 \Bigg.-\ln\left(\sqrt{h_{i}}+(4qh_{i}h_{f})^{1/4}\sqrt{\gamma_{1}}(-T)^{-\gamma_{0}/2}+\sqrt{qh_{f}}\gamma_{1}(-T)^{-\gamma_{0}}\right)\Bigg]\\
 \zeta_{2}(T,\mathcal{T})\equiv-\frac{\mathcal{T}\lambda a_{0}^{3(1+\omega)}\gamma_{1}^{3h_{i}(1+\omega)}(-T)^{-3h_{i}\gamma_{0}(1+\omega)}}{2\rho_{0}(1+\omega)}.
\end{eqnarray}
\item If $h_{i}<\frac{1}{3(1+\omega)}$, then
\begin{eqnarray}
%-\left(12k^{2}_{1}k_{0}+9k_{0}h_{i}^{3/2}(4qh_{i}h_{f})^{1/4}-3k_{0}h_{i}^{5/2}(4qh_{i}h_{f})^{1/4}\right)\Phi^{-3}\sim-\frac{4T\lambda}{3h_{f}}, 
\Phi\sim\gamma_{2}(-\frac{1}{T})^{1/3},
\end{eqnarray}
with 
\begin{eqnarray}
\gamma_{2}\equiv\left(\frac{9h_{f}k_{0}}{4\lambda}\right)^{1/3}\left(h^{5/2}_{i}(4qh_{i}h_{f})^{1/4}-4k_{1}^{2}-3h_{i}^{3/2}(4qh_{i}h_{f})^{1/1}\right)^{1/3}
\end{eqnarray}
Therefore 
%\begin{eqnarray}
% f(T,\mathcal{T})&\simeq&-C\sqrt{\gamma_{2}}(-T)^{5/6}-\frac{2\lambda\gamma_{2}^{2}(-T)^{1/3}}{3h_{f}}\nonumber\\
% &&-\left(6Ch^{2}_{i}k_{0}\gamma_{2}^{-3/2}(-T)^{1/2}+k_{0}\sqrt{\gamma_{2}}(-T)^{5/6}\right)\times\nonumber\\
% &&\left(Arctan\left(1-k_{1}\sqrt{\gamma_{2}}(-T)^{-1/6}\right)-Arctan\left(1+k_{1}\sqrt{\gamma_{2}}(-T)^{-1/6}\right)\right)\nonumber\\
% &&+\left(3k_{0}h^{2}_{i}\gamma_{2}^{-3/2}(-T)^{1/2}+\frac{k_{0}\sqrt{\gamma_{2}}(-T)^{5/6}}{2}\right)\times\nonumber\\
% &&\left(\ln\left(\sqrt{h_{i}}-(4qh_{i}h_{f})^{1/4}\sqrt{\gamma_{2}}(-T)^{-1/6}+\sqrt{qh_{f}}\gamma_{2}(-T)^{-1/3}\right)\right.\nonumber\\
%&& \left.-\ln\left(\sqrt{h_{i}}+(4qh_{i}h_{f})^{1/4}\sqrt{\gamma_{2}}(-T)^{-1/6}+\sqrt{qh_{f}}\gamma_{2}(-T)^{-1/3}\right)\right)\nonumber\\
%&&+\frac{\rho_{0}(1-5\omega)}{2a_{0}^{3(1+\omega)}\gamma_{2}^{3h_{i}(1+\omega)}}(-T)^{h_{i}(1+\omega)}-6Ch^{2}_{i}\gamma_{2}^{-3/2}(-T)^{1/2}\nonumber\\
%&&-\frac{\lambda a_{0}^{3(1+\omega)}\mathcal{T}\gamma_{2}^{3h_{i}(1+\omega)}}{2\rho_{0}(1+\omega)(-T)^{h_{i}(1+\omega)}}-\frac{1}{2}\mathcal{T}.
%\end{eqnarray}
\begin{eqnarray}
 f(T,\mathcal{T})&\simeq&\zeta_{3}(T)+ K_{2}(-T)^{5/6}+K_{3}(-T)^{1/3}+K_{4}(-T)^{1/2}\nonumber\\
 &&+K_{5}(-T)^{h_{i}(1+\omega)}+\frac{K_{6}\mathcal{T}}{(-T)^{h_{i}(1+\omega)}}-\frac{1}{2}\mathcal{T},
\end{eqnarray}
where 
\begin{eqnarray}
 \zeta_{3}(T)\equiv-\Bigg[6P_{0}h^{2}_{i}k_{0}\gamma_{2}^{-3/2}(-T)^{1/2}+k_{0}\sqrt{\gamma_{2}}(-T)^{5/6}\Bigg]\times\nonumber\\
 \Bigg[Arctan\left(1-k_{1}\sqrt{\gamma_{2}}(-T)^{-1/6}\right)-Arctan\left(1+k_{1}\sqrt{\gamma_{2}}(-T)^{-1/6}\right)\Bigg]\nonumber\\
 +\Bigg[3k_{0}h^{2}_{i}\gamma_{2}^{-3/2}(-T)^{1/2}+\frac{k_{0}\sqrt{\gamma_{2}}(-T)^{5/6}}{2}\Bigg]\times\nonumber\\
\Bigg[\ln\left(\sqrt{h_{i}}-(4qh_{i}h_{f})^{1/4}\sqrt{\gamma_{2}}(-T)^{-1/6}+\sqrt{qh_{f}}\gamma_{2}(-T)^{-1/3}\right)\Bigg.\nonumber\\
 \Bigg.-\ln\left(\sqrt{h_{i}}+(4qh_{i}h_{f})^{1/4}\sqrt{\gamma_{2}}(-T)^{-1/6}+\sqrt{qh_{f}}\gamma_{2}(-T)^{-1/3}\right)\Bigg],
\end{eqnarray}
\begin{eqnarray}
 K_{2}\equiv-P_{0}\sqrt{\gamma_{2}},K_{3}\equiv-\frac{2\lambda\gamma_{2}^{2}}{3h_{f}}, K_{4}\equiv-6P_{0}h^{2}_{i}\gamma_{2}^{-3/2},\\
 K_{5}\equiv\frac{\rho_{0}(1-5\omega)}{2a_{0}^{3(1+\omega)}\gamma_{2}^{3h_{i}(1+\omega)}},K_{6}\equiv-\frac{\lambda a_{0}^{3(1+\omega)}\gamma_{2}^{3h_{i}(1+\omega)}}{2\rho_{0}(1+\omega)}.
\end{eqnarray}
\end{itemize}
\item When $\Phi\longrightarrow+\infty$, one gets 
\begin{eqnarray}
 P_{2}(\Phi)\simeq-\frac{1}{2}\left(1+\frac{\lambda a_{0}^{3(1+\omega)}}{\rho_{0}(1+\omega)}\Phi^{3h_{f}(1+\omega)}\right),
\end{eqnarray}
and 
\begin{eqnarray}
 Q(\Phi)\simeq\frac{\rho_{0}(1-5\omega)}{2a_{0}^{3(1+\omega)}}\Phi^{-3h_{f}(1+\omega)}-6P_{0}h^{2}_{f}\Phi^{-3/2}\nonumber\\
 -6P_{0}h^{2}_{f}k_{0}\Phi^{-3/2}\Bigg[Arctan(1-k_{1}\sqrt{\Phi})-Arctan(1+k_{1}\sqrt{\Phi})\Bigg]\nonumber\\
 +3k_{0}h^{2}_{f}\Phi^{-3/2}\Bigg[\ln\left(\sqrt{h_{i}}-(4qh_{i}h_{f})^{1/4}\sqrt{\Phi}+\sqrt{qh_{f}}\Phi\right)\Bigg.\nonumber\\
\Bigg. -\ln\left(\sqrt{h_{i}}+(4qh_{i}h_{f})^{1/4}\sqrt{\Phi}+\sqrt{qh_{f}}\Phi\right)\Bigg].
\end{eqnarray}
From the relation (\ref{b}), it appears that
\begin{eqnarray}
 \frac{4T\lambda}{3h_{f}}\Phi^{4}+\frac{P_{0}T}{2}\Phi^{5/2}-\left(\frac{k_{0}T}{k_{1}}-\frac{k_{1}T(4qh_{i}h_{f})^{1/4}}{2\sqrt{qh_{f}}}\right)\Phi^{2}\nonumber\\
 +\left(9P_{0}h_{f}^{2}\Phi^{1/2}-\frac{k_{0}T}{2}\Phi^{5/2}\right)\Bigg[Arctan(1-k_{1}\sqrt{\Phi})-Arctan(1+k_{1}\sqrt{\Phi})\Bigg]\nonumber\\
 9P_{0}h_{f}^{2}\Phi^{1/2}-\frac{3h_{f}\lambda\mathcal{T}a_{0}^{3(1+\omega)}}{2\rho_{0}}\Phi^{3h_{f}(1+\omega)+5/2}-\frac{3h_{f}(1+\omega)(1-5\omega)\rho_{0}}{2a_{0}^{3(1+\omega)}}\Phi^{-3h_{f}(1+\omega)+5/2}\nonumber\\
 +\left(\frac{3k_{0}k_{1}^{2}(4qh_{i}h_{f})^{1/4}}{\sqrt{qh_{f}}}+\frac{6P_{0}h_{f}^{2}k_{0}}{k_{1}^{2}}\right)\sim0.
\end{eqnarray}
Then, we distinguish the following cases 
\begin{itemize}
 \item If $3h_{f}(1+\omega)+5/2>4\Rightarrow h_{f}>\frac{1}{2(1+\omega)}$, one gets
% \begin{eqnarray}
%  \frac{3h_{f}\lambda\mathcal{T}a_{0}^{3(1+\omega)}}{2\rho_{0}}\Phi^{3h_{f}(1+\omega)+5/2}\sim\left(\frac{3k_{0}k_{1}^{2}(4qh_{i}h_{f})^{1/4}}{\sqrt{qh_{f}}}+\frac{6Ch_{f}^{2}k_{0}}{k_{1}^{2}}\right),\nonumber\\
%  \Phi^{3h_{f}(1+\omega)+5/2}\sim\frac{2\rho_{0}}{3h_{f}\lambda\mathcal{T}a_{0}^{3(1+\omega)}}\left(\frac{3k_{0}k_{1}^{2}(4qh_{i}h_{f})^{1/4}}{\sqrt{qh_{f}}}+\frac{6Ch_{f}^{2}k_{0}}{k_{1}^{2}}\right),\nonumber\\
%  \Phi\sim\left(\frac{2\rho_{0}}{3h_{f}\lambda\mathcal{T}a_{0}^{3(1+\omega)}}\right)^{\frac{2}{6h_{i}(1+\omega)+5}}\left(\frac{3k_{0}k_{1}^{2}(4qh_{i}h_{f})^{1/4}}{\sqrt{qh_{f}}}+\frac{6Ch_{f}^{2}k_{0}}{k_{1}^{2}}\right)^{\frac{2}{6h_{i}(1+\omega)+5}}
% \end{eqnarray}
 \begin{eqnarray}
  \Phi\sim\delta_{1}\mathcal{T}^{-\delta_{0}},
 \end{eqnarray}
 with 
 \begin{eqnarray}
  \delta_{0}\equiv\frac{2}{6h_{i}(1+\omega)+5},\\
  \delta_{1}\equiv\left(\frac{2\rho_{0}}{3h_{f}\lambda a_{0}^{3(1+\omega)}}\right)^{\frac{2}{6h_{i}(1+\omega)+5}}\left(\frac{3k_{0}k_{1}^{2}(4qh_{i}h_{f})^{1/4}}{\sqrt{qh_{f}}}+\frac{6Ch_{f}^{2}k_{0}}{k_{1}^{2}}\right)^{\frac{2}{6h_{i}(1+\omega)+5}}
 \end{eqnarray}
 This yields  
% \begin{eqnarray}
%  f(T,\mathcal{T})&\simeq&C\sqrt{\delta_{1}}T\mathcal{T}^{-\delta_{0}/2}+\frac{2\lambda\delta_{1}^{2}T\mathcal{T}^{-2\delta_{0}}}{3h_{f}}\nonumber\\
%  &&+\left(k_{0}\sqrt{\delta_{1}}T\mathcal{T}^{-\delta_{0}/2}-6Ch^{2}_{f}k_{0}\delta_{1}^{-3/2}\mathcal{T}^{3\delta_{0}/2}\right)\times\\
%  &&\left(Arctan\left(1-k_{1}\sqrt{\delta_{1}}\mathcal{T}^{-\delta_{0}/2}\right)-Arctan\left(1+k_{1}\sqrt{\delta_{1}}\mathcal{T}^{-\delta_{0}/2}\right)\right)\nonumber\\
% &&+\left(3k_{0}h^{2}_{f}\delta_{1}^{-3/2}\mathcal{T}^{3\delta_{0}/2}-\frac{k_{0}\sqrt{\delta_{1}}T\mathcal{T}^{-\delta_{0}/2}}{2}\right)\times\\
% &&\left(\ln\left(\sqrt{h_{i}}-(4qh_{i}h_{f})^{1/4}\sqrt{\delta_{1}}\mathcal{T}^{-\delta_{0}/2}+\sqrt{qh_{f}}\delta_{1}\mathcal{T}^{-\delta_{0}}\right)\right.\nonumber\\
%&& \left.-\ln\left(\sqrt{h_{i}}+(4qh_{i}h_{f})^{1/4}\sqrt{\delta_{1}}\mathcal{T}^{-\delta_{0}/2}+\sqrt{qh_{f}}\delta_{1}\mathcal{T}^{-\delta_{0}}\right)\right)\nonumber\\
%&&\frac{\rho_{0}(1-5\omega)}{2a_{0}^{3(1+\omega)}\delta_{1}^{3h_{f}(1+\omega)}}\mathcal{T}^{3\delta_{0}h_{f}(1+\omega)}-6Ch^{2}_{f}\delta_{1}^{-3/2}\mathcal{T}^{3\delta_{0}/2}\nonumber\\
%&&-\frac{\lambda a_{0}^{3(1+\omega)}\delta_{1}^{3h_{f}(1+\omega)}}{2\rho_{0}(1+\omega)}\mathcal{T}^{-3h_{f}\delta_{0}(1+\omega)}-\frac{1}{2}\mathcal{T}.
% \end{eqnarray}
 \begin{eqnarray}
  f(T,\mathcal{T})\simeq\zeta_{4}(T,\mathcal{T})+\zeta_{5}(T,\mathcal{T})+\zeta_{6}(\mathcal{T}),
 \end{eqnarray}
where
\begin{eqnarray}
 \zeta_{4}(T,\mathcal{T})\equiv P_{0}\sqrt{\delta_{1}}T\mathcal{T}^{-\delta_{0}/2}+\frac{2\lambda\delta_{1}^{2}T\mathcal{T}^{-2\delta_{0}}}{3h_{f}},\\
 \zeta_{5}(T,\mathcal{T})\equiv+\Bigg[k_{0}\sqrt{\delta_{1}}T\mathcal{T}^{-\delta_{0}/2}-6P_{0}h^{2}_{f}k_{0}\delta_{1}^{-3/2}\mathcal{T}^{3\delta_{0}/2}\Bigg]\times\\
 \Bigg[Arctan\left(1-k_{1}\sqrt{\delta_{1}}\mathcal{T}^{-\delta_{0}/2}\right)-Arctan\left(1+k_{1}\sqrt{\delta_{1}}\mathcal{T}^{-\delta_{0}/2}\right)\Bigg]\nonumber\\
+\Bigg[3k_{0}h^{2}_{f}\delta_{1}^{-3/2}\mathcal{T}^{3\delta_{0}/2}-\frac{k_{0}\sqrt{\delta_{1}}T\mathcal{T}^{-\delta_{0}/2}}{2}\Bigg]\times\\
\Bigg[\ln\left(\sqrt{h_{i}}-(4qh_{i}h_{f})^{1/4}\sqrt{\delta_{1}}\mathcal{T}^{-\delta_{0}/2}+\sqrt{qh_{f}}\delta_{1}\mathcal{T}^{-\delta_{0}}\right)\Bigg.\nonumber\\
 \Bigg.-\ln\left(\sqrt{h_{i}}+(4qh_{i}h_{f})^{1/4}\sqrt{\delta_{1}}\mathcal{T}^{-\delta_{0}/2}+\sqrt{qh_{f}}\delta_{1}\mathcal{T}^{-\delta_{0}}\right)\Bigg]\\
 \zeta_{6}(\mathcal{T})\equiv\frac{\rho_{0}(1-5\omega)}{2a_{0}^{3(1+\omega)}\delta_{1}^{3h_{f}(1+\omega)}}\mathcal{T}^{3\delta_{0}h_{f}(1+\omega)}-6P_{0}h^{2}_{f}\delta_{1}^{-3/2}\mathcal{T}^{3\delta_{0}/2}\nonumber\\
-\frac{\lambda a_{0}^{3(1+\omega)}\delta_{1}^{3h_{f}(1+\omega)}}{2\rho_{0}(1+\omega)}\mathcal{T}^{-3h_{f}\delta_{0}(1+\omega)}-\frac{1}{2}\mathcal{T}
\end{eqnarray}
 \item If $3h_{f}(1+\omega)+5/2<4\Rightarrow h_{f}<\frac{1}{2(1+\omega)}$, then 
 \begin{eqnarray}
%  \frac{4T\lambda}{3h_{f}}\Phi^{4}\sim-\left(\frac{3k_{0}k_{1}^{2}(4qh_{i}h_{f})^{1/4}}{\sqrt{qh_{f}}}+\frac{6Ch_{f}^{2}k_{0}}{k_{1}^{2}}\right),\nonumber\\
    \Phi\sim\left(-\frac{3h_{f}}{4T\lambda}\right)^{1/4}\left(\frac{3k_{0}k_{1}^{2}(4qh_{i}h_{f})^{1/4}}{\sqrt{qh_{f}}}+\frac{6P_{0}h_{f}^{2}k_{0}}{k_{1}^{2}}\right)^{1/4}\equiv\delta_{2}(-T)^{-1/4}.
 \end{eqnarray}
 %\begin{eqnarray}
% f(T,\mathcal{T})&=&-P_{0}\sqrt{\delta_{2}}(-T)^{7/8}-\frac{2\lambda\delta_{2}^{2}(-T)^{1/2}}{3h_{f}}\nonumber\\
% &&-\left(6P_{0}h^{2}_{f}k_{0}\delta_{2}^{-3/2}(-T)^{3/8}+k_{0}\sqrt{\delta_{2}}(-T)^{7/8}\right)\times\\
% &&\left(Arctan\left(1-k_{1}\sqrt{\delta_{2}}(-T)^{-1/8}\right)-Arctan\left(1+k_{1}\sqrt{\delta_{2}}(-T)^{-1/8}\right)\right)\nonumber\\
% &&+\left(3k_{0}h^{2}_{f}\delta_{2}^{-3/2}(-T)^{3/8}+\frac{k_{0}\sqrt{\delta_{2}}(-T)^{7/8}}{2}\right)\times\nonumber\\
% &&\left(\ln\left(\sqrt{h_{i}}-(4qh_{i}h_{f})^{1/4}\sqrt{\delta_{2}}(-T)^{-1/8}+\sqrt{qh_{f}}\delta_{2}(-T)^{-1/4}\right)\right.\nonumber\\
%&& \left.-\ln\left(\sqrt{h_{i}}+(4qh_{i}h_{f})^{1/4}\sqrt{\delta_{2}}(-T)^{-1/8}+\sqrt{qh_{f}}\delta_{2}(-T)^{-1/4}\right)\right)\nonumber\\
%&&+\frac{\rho_{0}(1-5\omega)}{2a_{0}^{3(1+\omega)}\delta_{2}^{3h_{f}(1+\omega)}}(-T)^{3h_{f}(1+\omega)/4}-6P_{0}h^{2}_{f}\delta_{2}^{-3/4}(-T)^{3/8}\nonumber\\
%&&-\frac{\lambda a_{0}^{3(1+\omega)}\delta_{2}^{3h_{f}(1+\omega)}}{2\rho_{0}(1+\omega)}\frac{\mathcal{T}}{(-T)^{3h_{f}(1+\omega)/4}}-\frac{1}{2}\mathcal{T},
%\end{eqnarray}
Then, one obtains 
\begin{eqnarray}
 f(T,\mathcal{T})&\equiv&\zeta_{7}(T)+K'_{0}(-T)^{7/8}+K'_{1}\sqrt{-T}+K'_{2}(-T)^{3/8}+K'_{3}(-T)^{3h_{f}(1+\omega)/4}\nonumber\\
 &&+\zeta_{8}(T,\mathcal{T})-\frac{1}{2}\mathcal{T},
\end{eqnarray}
with 
\begin{eqnarray}
 K'_{0}\equiv-P_{0}\sqrt{\delta_{2}},K'_{1}\equiv-\frac{2\lambda\delta_{2}^{2}}{3h_{f}},K'_{2}\equiv-6P_{0}h^{2}_{f}\delta_{2}^{-3/4},\\
 K'_{3}\equiv\frac{\rho_{0}(1-5\omega)}{2a_{0}^{3(1+\omega)}\delta_{2}^{3h_{f}(1+\omega)}},
\end{eqnarray}
and 
\begin{eqnarray}
 \zeta_{7}(T)\equiv-\Bigg[6P_{0}h^{2}_{f}k_{0}\delta_{2}^{-3/2}(-T)^{3/8}+k_{0}\sqrt{\delta_{2}}(-T)^{7/8}\Bigg]\times\nonumber\\
\Bigg[Arctan\left(1-k_{1}\sqrt{\delta_{2}}(-T)^{-1/8}\right)-Arctan\left(1+k_{1}\sqrt{\delta_{2}}(-T)^{-1/8}\right)\Bigg]\nonumber\\
+\Bigg[3k_{0}h^{2}_{f}\delta_{2}^{-3/2}(-T)^{3/8}+\frac{k_{0}\sqrt{\delta_{2}}(-T)^{7/8}}{2}\Bigg]\times\nonumber\\
\Bigg[\ln\left(\sqrt{h_{i}}-(4qh_{i}h_{f})^{1/4}\sqrt{\delta_{2}}(-T)^{-1/8}+\sqrt{qh_{f}}\delta_{2}(-T)^{-1/4}\right)\Bigg.\nonumber\\
\Bigg.-\ln\left(\sqrt{h_{i}}+(4qh_{i}h_{f})^{1/4}\sqrt{\delta_{2}}(-T)^{-1/8}+\sqrt{qh_{f}}\delta_{2}(-T)^{-1/4}\right)\Bigg],\\
\zeta_{8}(T,\mathcal{T})\equiv-\frac{\lambda a_{0}^{3(1+\omega)}\delta_{2}^{3h_{f}(1+\omega)}}{2\rho_{0}(1+\omega)}\frac{\mathcal{T}}{(-T)^{3h_{f}(1+\omega)/4}}.
\end{eqnarray}
\end{itemize}
\end{enumerate}
%%%%%%%%%%%%%%%%%%%%%%%%%%%%%%%%%%%%%%%%%%%%%%%%%%%%%%%%%%%%%%%%%%%%%%%%%%%%%%%%%%%%%%%%%%%%%%%%%%%%%%%%%%%%%%%%%%%%%%%%%%%%%%%%%%%%%%%%%%%%%%%%%%%%%%%%%%%%%%%%%%%%%%%%%%%%%%%%%%%%%%%%%%%%%%%%%%%%%%%%%%%%%%%%%%%%%%%%%%%%%%%%%%%%%%%%%%%%%%%%%%%%%%%%%%
\section{Conclusion}

The work developed in this paper concerns the $f(T,\mathcal{T})$ theory of gravity where $T$ and $\mathcal{T}$ denotes the torsion scalar and the trace of the energy-momentum tensor, respectively. It is well known from the observational data that in order to explain the interaction between the matter and dark energy it necessary to the cosmological constant be variable and may therefore dependent on the trace of the energy-momentum tensor. After the inflation, the universe has been dominated by the matter where its expansion were decelerated and now it is dominated by the dark energy with an accelerated expansion. We search for $f(T,\mathcal{T})$ models able to describe the matter and dark energy dominated phases and their unification. To do so, we consider two cosmological expressions for the scale factor and our results show that it is well possible to $f(T,\mathcal{T})$ models describe the different phases of the expansion of the universe, i.e, the matter dominated phase and the dark energy dominated phase. \par 
More precisely, we distinguish two interesting cases, a particular case and a more general case through an input parameter $\lambda$. We observe that when this parameter vanishes, the results of the particular case are recovered. We essential part of the calculus is based on the expression of the scale factor. Then, we observe two fundamental expressions which are cosmologically realistic. The first concerns the form  (\ref{19}) with the condition (\ref{18}) within what we obtain the algebraic expressions of the gravitational action, one characterising early times for small $\Phi$ and the second, the late time accelerated phase for large $\Phi$. Moreover, our results point out that at early time, the universe may effectively be dominated by both relativistic and non-relativistic matter. On the other hand, we also see that for large times characterising the present phase, the universe is essential filled by the dark energy, proving the consistency of our results. The second form  of the  scale factor still obeys (\ref{19}) but in this case with (\ref{24}). Here, we use the adiabatic form of the expression of the scale factor, our results show the transition from the matter dominated  phase to the late time dark energy dominated phase.\par 
However, an interesting aspect to be undertaken in this paper is studying the stability of such models and their thermodynamics. We leave this as a future work.

\vspace{0.5cm}
{\bf Acknowledgement}: SB Nassur thanks DAAD for financial support. 
MJS Houndjo, VA Kpadonou and J Tossa would like to thank {\it Ecole Normale Sup\'erieure Natitingou for partial financial backing during the elaboration of this work. ME Rodrigues thanks UFPA, Edital 04/2014 PROPESP, and CNPq Edital MCTI / CNPQ / Universal 14/2014, for partial financial support.

%%%%%%%%%%%%%%%%%%%%%%%%%%%%%%%%%%%%%%%%%%%%%%%%%%%%%%%%%%%%%%%%%%%%%%%%%%%%%%%%%%%%%%%%%%%%%%%%%%%%%%%%%%%%%%%%%%%%%%%%%%%%%%%%%

%\maketitle

%\section{}

\end{document}